\documentclass[preprint2]{aastex6}

\usepackage{booktabs}
\usepackage{longtable}
\usepackage{xspace}
\usepackage[T1]{fontenc}

\newcommand{\Kepler}{\textit{Kepler}\xspace} 
\newcommand{\spitzer}{\textit{Spitzer}\xspace} 
\newcommand{\ktwo}{\textit{K2}\xspace}

\newcommand{\Mstar}{\ensuremath{M_{\star}}\xspace}
\newcommand{\Rstar}{\ensuremath{R_{\star}}\xspace} 
 
\newcommand{\fe}{[Fe/H]\xspace}
\newcommand{\teff}{\ensuremath{T_{\mathrm{eff}}}\xspace}  
\newcommand{\logg}{\ensuremath{\log g}\xspace}

\newcommand{\Mp}{\ensuremath{M_{p}}\xspace} 
\newcommand{\Rp}{\ensuremath{R_p}\xspace}

\newcommand{\fenv}{\ensuremath{f_{\mathrm{env}}}\xspace}
\newcommand{\menv}{\ensuremath{M_{\mathrm{env}}}\xspace}
\newcommand{\mcore}{\ensuremath{M_{\mathrm{core}}}\xspace}

\newcommand{\ms}{m s$^{-1}$\xspace}

\newcommand{\Me}{\ensuremath{M_{\oplus}}\xspace} 
\renewcommand{\Re}{\ensuremath{R_{\oplus}}\xspace} 
\newcommand{\um}{\ensuremath{\mu \mathrm{m}}\xspace}

\newcommand{\Rsun}{\ensuremath{R_{\odot}}\xspace }
\newcommand{\Msun}{\ensuremath{M_{\odot}}\xspace}

\def\deg{\ensuremath{^{\circ}}}

\newcommand{\K}[1]{\ensuremath{K_\mathrm{#1}}\xspace}

\newcommand{\sigjit}[1]{
        \ensuremath{
                \ifthenelse{\equal{#1}{}}{\sigma_\mathrm{jit}}{\sigma_\mathrm{jit,#1}}}
        \xspace
}
\usepackage{xstring} 
\usepackage{ifthen} 
\usepackage{xifthen} 

\newcommand{\argperi}[1]{\ensuremath{\ifthenelse{\isempty{#1}}{\omega}{\omega_{#1}}}\xspace}
\newcommand{\inc}[1]{\ensuremath{\ifthenelse{\isempty{#1}}{i}{i_{#1}}}\xspace}
\newcommand{\ecc}[1]{\ensuremath{\ifthenelse{\isempty{#1}}{e}{e_{#1}}}\xspace}
\newcommand{\per}[1]{\ensuremath{\ifthenelse{\isempty{#1}}{P}{P_{#1}}}\xspace}
\newcommand{\node}[1]{\ensuremath{\ifthenelse{\isempty{#1}}{\Omega}{\Omega_{#1}}}\xspace}
\newcommand{\meananom}[1]{\ensuremath{\ifthenelse{\isempty{#1}}{M}{M_{#1}}}\xspace}
\newcommand{\ecosw}[1]{\ensuremath{\ifthenelse{\isempty{#1}}{e \cos \omega}{e \cos \omega_{#1}}}\xspace}
\newcommand{\esinw}[1]{\ensuremath{\ifthenelse{\isempty{#1}}{e \sin \omega}{e \sin \omega_{#1}}}\xspace}
\newcommand{\secosw}[1]{\ensuremath{\sqrt{e_{#1}} \cos \omega_{#1}}\xspace}
\newcommand{\sesinw}[1]{\ensuremath{\sqrt{e_{#1}} \sin \omega_{#1}}\xspace}

\newcommand{\meanecc}{\ensuremath{\langle e \rangle}\xspace}

\newcommand{\T}[1]{\ensuremath{\ifthenelse{\isempty{#1}}{T}{T_{#1}}}\xspace}
\newcommand{\tc}[1]{\ensuremath{\ifthenelse{\isempty{#1}}{T_{c}}{T_{c,#1}}}\xspace}

\newcommand{\Zfree}{\ensuremath{Z_{\mathrm{free}}}\xspace} 
\newcommand{\Zfreestar}{\ensuremath{Z_{\mathrm{free}}^{*}}\xspace}
\newcommand{\zfree}[1]{\ensuremath{\ifthenelse{\isempty{#1}}{z_{\mathrm{free}}^{*}}{z_{\mathrm{free,#1}}}}\xspace}

\renewcommand{\Im}{\ensuremath{\mathrm{Im}}}

\newcommand{\ReZfree}{\ensuremath{\mathrm{Re}(\Zfree)}}
\newcommand{\ImZfree}{\ensuremath{\Im(\Zfree)}}
\newcommand{\mr}{\ensuremath{ M_{p,2} / M_{p,1}}\xspace}

\newcommand{\stellar}[1]{%
\IfEqCase{#1}{%
{br-teff}{ $5625 \pm 60$ }%
{br-logg}{ $4.29 \pm 0.05$ }%
{br-fe}{ $+0.34 \pm 0.04$ }%
{br-mass}{ $1.07 \pm 0.06$ }%
{br-radius}{ $1.16 \pm 0.04$ }%
}[\PackageError{tree}{Undefined option to tree: #1}{}]%
}%

\newcommand{\stat}[1]{%
\IfEqCase{#1}{%
{p16-ror1}{4.31}
{p16-ror1_err}{0.12}
{p16-ror2}{5.94}
{p16-ror2_err}{0.07}
{rv-n}{63}
{rv-start}{2015-06-24}
{rv-stop}{2017-10-03}
{rv-errvel-min}{1.5}
{rv-errvel-max}{2.1}
{ref-ephem-per-1}{20.889939}
{ref-ephem-tc-1}{2072.889}
{ref-ephem-per-2}{42.34066}
{ref-ephem-tc-2}{2082.423}
{lithwick-prior-mu1}{48}
{lithwick-prior-mu1_err}{9}
{lithwick-prior-mu2}{53}
{lithwick-prior-mu2_err}{11}
{ttvfast-tstart}{2065}
{ttvfast-emcee-nwalkers}{1000}
{ttvfast-emcee-nsteps}{2000}
{ttvfast-emcee-nburn}{500}
}[\PackageError{tree}{Undefined option to tree: #1}{}]%
}%
\newcommand{\hand}[1]{%
\IfEqCase{#1}{%
{fit-ccc-bic}{366.0}
{fit-cc-bic}{378.6}
{fit-cce-bic}{381.2}
{masse1-p16_fmt}{$21.0 \pm 5.4$}
{masse2-p16_fmt}{$27.0 \pm 6.9$}
{fenv1-p17_fmt}{$28^{+7}_{-6}$}
{fenv2-p17_fmt}{$57^{+9}_{-10}$}
{mcore2-p17_fmt}{$11.5^{+3.0}_{-3.0}$}
{mcore1-p17_fmt}{$15.0^{+3.9}_{-3.9}$}
}[\PackageError{tree}{Undefined option to tree: #1}{}]%
}%

\newcommand{\fit}[1]{%
\IfEqCase{#1}{%
{ccc-per1}{20.88526}
{ccc-per1_err1}{0.00000}
{ccc-per1_err2}{0.00000}
{ccc-per1_fmt}{$20.88526^{+0.00000}_{0.00000}$}
{ccc-per2}{42.36301}
{ccc-per2_err1}{0.00000}
{ccc-per2_err2}{0.00000}
{ccc-per2_fmt}{$42.36301^{+0.00000}_{0.00000}$}
{ccc-per3}{437}
{ccc-per3_err1}{32}
{ccc-per3_err2}{-25}
{ccc-per3_fmt}{$437^{+32}_{-25}$}
{ccc-tc1}{2072.79}
{ccc-tc1_err1}{0.00}
{ccc-tc1_err2}{0.00}
{ccc-tc1_fmt}{$2072.79^{+0.00}_{0.00}$}
{ccc-tc2}{2082.63}
{ccc-tc2_err1}{0.00}
{ccc-tc2_err2}{0.00}
{ccc-tc2_fmt}{$2082.63^{+0.00}_{0.00}$}
{ccc-tc3}{2374}
{ccc-tc3_err1}{27}
{ccc-tc3_err2}{-29}
{ccc-tc3_fmt}{$2374^{+27}_{-29}$}
{ccc-k1}{3.8}
{ccc-k1_err1}{0.7}
{ccc-k1_err2}{-0.7}
{ccc-k1_fmt}{$3.8^{+0.7}_{-0.7}$}
{ccc-k2}{3.4}
{ccc-k2_err1}{0.7}
{ccc-k2_err2}{-0.7}
{ccc-k2_fmt}{$3.4^{+0.7}_{-0.7}$}
{ccc-k3}{4.4}
{ccc-k3_err1}{1.1}
{ccc-k3_err2}{-1.2}
{ccc-k3_fmt}{$4.4^{+1.1}_{-1.2}$}
{ccc-a1}{0.152}
{ccc-a1_err1}{0.003}
{ccc-a1_err2}{-0.003}
{ccc-a1_fmt}{$0.152^{+0.003}_{-0.003}$}
{ccc-a2}{0.243}
{ccc-a2_err1}{0.004}
{ccc-a2_err2}{-0.005}
{ccc-a2_fmt}{$0.243^{+0.004}_{-0.005}$}
{ccc-a3}{1.15}
{ccc-a3_err1}{0.06}
{ccc-a3_err2}{-0.05}
{ccc-a3_fmt}{$1.15^{+0.06}_{-0.05}$}
{ccc-mpsini1}{16.8}
{ccc-mpsini1_err1}{3.2}
{ccc-mpsini1_err2}{-3.1}
{ccc-mpsini1_fmt}{$16.8^{+3.2}_{-3.1}$}
{ccc-mpsini2}{19.0}
{ccc-mpsini2_err1}{3.9}
{ccc-mpsini2_err2}{-3.8}
{ccc-mpsini2_fmt}{$19.0^{+3.9}_{-3.8}$}
{ccc-mpsini3}{54}
{ccc-mpsini3_err1}{14}
{ccc-mpsini3_err2}{-14}
{ccc-mpsini3_fmt}{$54^{+14}_{-14}$}
{ccc-musini1}{48.2}
{ccc-musini1_err1}{9.0}
{ccc-musini1_err2}{-8.8}
{ccc-musini1_fmt}{$48.2^{+9.0}_{-8.8}$}
{ccc-musini2}{54.4}
{ccc-musini2_err1}{11.0}
{ccc-musini2_err2}{-10.8}
{ccc-musini2_fmt}{$54.4^{+11.0}_{-10.8}$}
{ccc-musini3}{155.3}
{ccc-musini3_err1}{39.5}
{ccc-musini3_err2}{-40.9}
{ccc-musini3_fmt}{$155.3^{+39.5}_{-40.9}$}
{ccc-gamma_j}{-0.2}
{ccc-gamma_j_err1}{0.9}
{ccc-gamma_j_err2}{-0.9}
{ccc-gamma_j_fmt}{$-0.2^{+0.9}_{-0.9}$}
{ccc-jit_j}{3.3}
{ccc-jit_j_err1}{0.4}
{ccc-jit_j_err2}{-0.4}
{ccc-jit_j_fmt}{$3.3^{+0.4}_{-0.4}$}
{ccc-dvdt}{1.6}
{ccc-dvdt_err1}{1.2}
{ccc-dvdt_err2}{-1.1}
{ccc-dvdt_fmt}{$1.6^{+1.2}_{-1.1}$}
{eec-e1_p90}{0.39}
{eec-e2_p90}{0.34}
}[\PackageError{tree}{Undefined option to tree: #1}{}]%
}%
\newcommand{\ttvfast}[1]{%
\IfEqCase{#1}{%
{muppm1}{53.2}
{muppm1_err1}{11.6}
{muppm1_err2}{-9.8}
{muppm1_fmt}{$53.2^{+11.6}_{-9.8}$}
{per1}{20.8850}
{per1_err1}{0.0004}
{per1_err2}{-0.0004}
{per1_fmt}{$20.8850^{+0.0004}_{-0.0004}$}
{secosw1}{0.24}
{secosw1_err1}{0.04}
{secosw1_err2}{-0.04}
{secosw1_fmt}{$0.24^{+0.04}_{-0.04}$}
{sesinw1}{-0.09}
{sesinw1_err1}{0.04}
{sesinw1_err2}{-0.05}
{sesinw1_fmt}{$-0.09^{+0.04}_{-0.05}$}
{tc1}{2072.7961}
{tc1_err1}{0.0008}
{tc1_err2}{-0.0008}
{tc1_fmt}{$2072.7961^{+0.0008}_{-0.0008}$}
{muppm2}{42.2}
{muppm2_err1}{10.3}
{muppm2_err2}{-8.6}
{muppm2_fmt}{$42.2^{+10.3}_{-8.6}$}
{per2}{42.3688}
{per2_err1}{0.0013}
{per2_err2}{-0.0012}
{per2_fmt}{$42.3688^{+0.0013}_{-0.0012}$}
{secosw2}{0.01}
{secosw2_err1}{0.10}
{secosw2_err2}{-0.10}
{secosw2_fmt}{$0.01^{+0.10}_{-0.10}$}
{sesinw2}{-0.03}
{sesinw2_err1}{0.11}
{sesinw2_err2}{-0.10}
{sesinw2_fmt}{$-0.03^{+0.11}_{-0.10}$}
{tc2}{2082.6261}
{tc2_err1}{0.0006}
{tc2_err2}{-0.0006}
{tc2_fmt}{$2082.6261^{+0.0006}_{-0.0006}$}
{masse1}{18.6}
{masse1_err1}{4.1}
{masse1_err2}{-3.5}
{masse1_fmt}{$18.6^{+4.1}_{-3.5}$}
{masse2}{14.7}
{masse2_err1}{3.6}
{masse2_err2}{-3.0}
{masse2_fmt}{$14.7^{+3.6}_{-3.0}$}
{mr2}{0.79}
{mr2_err1}{0.03}
{mr2_err2}{-0.03}
{mr2_fmt}{$0.79^{+0.03}_{-0.03}$}
{e1}{0.07}
{e1_err1}{0.01}
{e1_err2}{-0.01}
{e1_fmt}{$0.07^{+0.01}_{-0.01}$}
{e2}{0.02}
{e2_err1}{0.02}
{e2_err2}{-0.01}
{e2_fmt}{$0.02^{+0.02}_{-0.01}$}
{prad1}{5.7}
{prad1_err1}{0.4}
{prad1_err2}{-0.4}
{prad1_fmt}{$5.7^{+0.4}_{-0.4}$}
{prad2}{7.8}
{prad2_err1}{0.5}
{prad2_err2}{-0.5}
{prad2_fmt}{$7.8^{+0.5}_{-0.5}$}
{rho1}{0.56}
{rho1_err1}{0.19}
{rho1_err2}{-0.14}
{rho1_fmt}{$0.56^{+0.19}_{-0.14}$}
{rho2}{0.17}
{rho2_err1}{0.06}
{rho2_err2}{-0.04}
{rho2_fmt}{$0.17^{+0.06}_{-0.04}$}
{e1_p90}{0.09}
{e2_p90}{0.04}
{fenv1}{26}
{fenv1_err1}{3}
{fenv1_err2}{-3}
{fenv1_fmt}{$26^{+3}_{-3}$}
{fenv2}{52}
{fenv2_err1}{5}
{fenv2_err2}{-3}
{fenv2_fmt}{$52^{+5}_{-3}$}
{mcore1}{14.0}
{mcore1_err1}{1.6}
{mcore1_err2}{-1.7}
{mcore1_fmt}{$14.0^{+1.6}_{-1.7}$}
{mcore2}{7.4}
{mcore2_err1}{1.1}
{mcore2_err2}{-1.0}
{mcore2_fmt}{$7.4^{+1.1}_{-1.0}$}
{menv1}{4.9}
{menv1_err1}{0.8}
{menv1_err2}{-0.7}
{menv1_fmt}{$4.9^{+0.8}_{-0.7}$}
{menv2}{8.0}
{menv2_err1}{1.4}
{menv2_err2}{-1.1}
{menv2_fmt}{$8.0^{+1.4}_{-1.1}$}
}[\PackageError{tree}{Undefined option to tree: #1}{}]%
}%
\newcommand{\lithwick}[1]{%
\IfEqCase{#1}{%
{per1}{20.88977}
{per1_err1}{0.00034}
{per1_err2}{-0.00035}
{per1_fmt}{$20.88977^{+0.00034}_{-0.00035}$}
{tc1}{2072.8855}
{tc1_err1}{0.0055}
{tc1_err2}{-0.0053}
{tc1_fmt}{$2072.8855^{+0.0055}_{-0.0053}$}
{muppm1}{53.3}
{muppm1_err1}{5.2}
{muppm1_err2}{-5.2}
{muppm1_fmt}{$53.3^{+5.2}_{-5.2}$}
{per2}{42.3391}
{per2_err1}{0.0012}
{per2_err2}{-0.0012}
{per2_fmt}{$42.3391^{+0.0012}_{-0.0012}$}
{tc2}{2082.4485}
{tc2_err1}{0.0078}
{tc2_err2}{-0.0079}
{tc2_fmt}{$2082.4485^{+0.0078}_{-0.0079}$}
{muppm2}{43.4}
{muppm2_err1}{4.8}
{muppm2_err2}{-4.7}
{muppm2_fmt}{$43.4^{+4.8}_{-4.7}$}
{rezfree}{0.038}
{rezfree_err1}{0.004}
{rezfree_err2}{-0.003}
{rezfree_fmt}{$0.038^{+0.004}_{-0.003}$}
{imzfree}{0.070}
{imzfree_err1}{0.008}
{imzfree_err2}{-0.007}
{imzfree_fmt}{$0.070^{+0.008}_{-0.007}$}
{zmag}{0.080}
{zmag_err1}{0.009}
{zmag_err2}{-0.007}
{zmag_fmt}{$0.080^{+0.009}_{-0.007}$}
{mr2}{0.81}
{mr2_err1}{0.03}
{mr2_err2}{-0.02}
{mr2_fmt}{$0.81^{+0.03}_{-0.02}$}
{masse1}{19.0}
{masse1_err1}{2.2}
{masse1_err2}{-2.1}
{masse1_fmt}{$19.0^{+2.2}_{-2.1}$}
{masse2}{15.4}
{masse2_err1}{1.9}
{masse2_err2}{-1.8}
{masse2_fmt}{$15.4^{+1.9}_{-1.8}$}
{prad1}{5.4}
{prad1_err1}{0.2}
{prad1_err2}{-0.2}
{prad1_fmt}{$5.4^{+0.2}_{-0.2}$}
{prad2}{7.5}
{prad2_err1}{0.3}
{prad2_err2}{-0.3}
{prad2_fmt}{$7.5^{+0.3}_{-0.3}$}
{rho1}{0.64}
{rho1_err1}{0.12}
{rho1_err2}{-0.10}
{rho1_fmt}{$0.64^{+0.12}_{-0.10}$}
{rho2}{0.20}
{rho2_err1}{0.04}
{rho2_err2}{-0.03}
{rho2_fmt}{$0.20^{+0.04}_{-0.03}$}
{fenv1}{26}
{fenv1_err1}{3}
{fenv1_err2}{-3}
{fenv1_fmt}{$26^{+3}_{-3}$}
{fenv2}{52}
{fenv2_err1}{5}
{fenv2_err2}{-3}
{fenv2_fmt}{$52^{+5}_{-3}$}
{mcore1}{14.0}
{mcore1_err1}{1.6}
{mcore1_err2}{-1.7}
{mcore1_fmt}{$14.0^{+1.6}_{-1.7}$}
{mcore2}{7.4}
{mcore2_err1}{1.1}
{mcore2_err2}{-1.0}
{mcore2_fmt}{$7.4^{+1.1}_{-1.0}$}
{menv1}{4.9}
{menv1_err1}{0.8}
{menv1_err2}{-0.7}
{menv1_fmt}{$4.9^{+0.8}_{-0.7}$}
{menv2}{8.0}
{menv2_err1}{1.4}
{menv2_err2}{-1.1}
{menv2_fmt}{$8.0^{+1.4}_{-1.1}$}
{e1}{0.06}
{e1_err1}{0.01}
{e1_err2}{-0.01}
{e1_fmt}{$0.06^{+0.01}_{-0.01}$}
{e2}{0.04}
{e2_err1}{0.02}
{e2_err2}{-0.02}
{e2_fmt}{$0.04^{+0.02}_{-0.02}$}
{e1_p90}{0.07}
{e2_p90}{0.07}
}[\PackageError{tree}{Undefined option to tree: #1}{}]%
}%

\newcommand{\radvel}{\texttt{RadVel}\xspace}

\begin{document}

\title{Dynamics and Formation of the Near-Resonant K2-24 System: \\ Insights from Transit-Timing Variations and Radial Velocities}

\author{Erik A. Petigura\altaffilmark{1,2,8}}
\author{Bj\"orn Benneke\altaffilmark{1,3}} 
\author{Konstantin Batygin\altaffilmark{1}}
\author{Benjamin J. Fulton\altaffilmark{1,9}}
\author{Michael Werner\altaffilmark{4}} 
\author{Jessica E. Krick\altaffilmark{4}} 
\author{Varoujan Gorjian\altaffilmark{4}}
\author{Evan Sinukoff\altaffilmark{5,6}}
\author{Katherine M. Deck\altaffilmark{1}}
\author{Sean M. Mills\altaffilmark{1}}
\and
\author{Drake Deming\altaffilmark{7}} 

\altaffiltext{1}{Division of Geological and Planetary Sciences, California Institute of Technology, Pasadena, CA 91101, USA}
\altaffiltext{2}{petigura@caltech.edu}
\altaffiltext{3}{University of Montreal, Montreal, QC, H3T 1J4, Canada}
\altaffiltext{4}{Jet Propulsion Laboratory, California Institute of Technology, Pasadena, CA 91109, USA}
\altaffiltext{5}{Institute for Astronomy, University of Hawai`i at M\={a}noa, Honolulu, HI 96822, USA}
\altaffiltext{6}{Cahill Center for Astrophysics, California Institute of Technology, Pasadena, CA 91125, USA}
\altaffiltext{7}{Department of Astronomy, University of Maryland at College Park, College Park, MD, 20742, USA}
\altaffiltext{8}{Hubble Fellow}
\altaffiltext{9}{Texaco Fellow}

\begin{abstract}
While planets between the size of Uranus and Saturn are absent within the Solar System, the star K2-24 hosts two such planets, K2-24b and c, with radii equal to \lithwick{prad1}~\Re and \lithwick{prad2}~\Re, respectively. The two planets have orbital periods of 20.9~days and 42.4~days, residing only 1\% outside the nominal 2:1 mean-motion resonance. In this work, we present results from a coordinated observing campaign to measure planet masses and eccentricities that combines radial velocity (RV) measurements from Keck/HIRES and transit-timing measurements from \ktwo and {\em Spitzer}. K2-24b and c have low, but  non-zero, eccentricities of $e_1 \sim e_2 \sim 0.08$. The low observed eccentricities provide clues regarding the formation and dynamical evolution of K2-24b and K2-24c, suggesting that they could be the result of stochastic gravitational interactions with a turbulent protoplanetary disk, among other mechanisms. K2-24b and c are $19\pm2$~\Me and $15\pm2$~\Me, respectively; K2-24c is 20\% less massive than K2-24b, despite being 40\% larger. Their large sizes and low masses imply large envelope fractions, which we estimate at \lithwick{fenv1_fmt}\% and \lithwick{fenv2_fmt}\%. In particular, K2-24c's large envelope presents an intriguing challenge to the standard model of core nucleated accretion that predicts the onset of runaway accretion when \fenv $\approx$ 50\%.
\end{abstract}

\keywords{planets and satellites: individual (K2-24b,K2-24c) -- planets and satellites: dynamical evolution and stability -- planets and satellites: formation -- techniques: radial velocities -- techniques: photometric}

\section{Introduction}
The vast majority of our current understanding about the masses and orbits of extrasolar planets is based on two techniques: radial velocities (RVs) and transit-timing variations (TTVs). Typically, RVs constrain $\Mp \sin i$, the planet mass modulo an unknown inclination angle. For high signal-to-noise datasets, deviations from sinusoidal RV curves can reveal orbital eccentricities, and for a few exceptional systems, non-Keplerian orbital dynamics have been observed (see, e.g., GJ876; \citealt{Rivera10,Nelson16,Millholland18}). For transiting systems, the $\sin i$ ambiguity is negligible and RVs constrain planet mass and bulk composition directly. Such measurements have been made for planets as small as Earth (see, e.g., Kepler-78b; \citealt{Howard13,Pepe13}). Accordingly, RV mass measurements of transiting planets have helped reveal important trends in planetary bulk compositions, such as the onset of low density envelopes above $\Rp \approx 1.5~\Re$ \citep{Marcy14,Weiss14,Rogers15}.

While the early theoretical work on TTVs was developed a decade ago \citep{Agol05,Holman05}, TTVs were not observed until NASA's \Kepler mission provided high precision, long baseline photometry \citep{Holman10}.  The TTV technique has achieved some remarkable results such as precision mass measurements of small planets in the Kepler-36 system \citep{Carter12}, the discovery of a Laplace-like resonance in the Kepler-223 system \citep{Mills16}, and mass measurements of non-transiting planets in the Kepler-88 system \citep{Nesvorny13}.

While the RV and TTV techniques have been applied to many individual systems, only a handful of systems have benefited from joint analyses. Systems with TTVs have almost exclusively been discovered during the prime \Kepler mission (\citealt{Borucki10a}; 2009--2013), which surveyed only 1/400 of the sky. While $\approx$40\% of \Kepler planets are in multi-planet systems \citep{Rowe14}, planets typically need to be near mean-motion resonance to produce detectable TTVs. \cite{Holczer16} reported TTVs for $\approx$260 \Kepler planets, but most are too faint for precision RV measurements with current-generation instruments, which typically require host stars with $V \lesssim 13$~mag. As a result, fewer than 10 systems have mass constraints from both the TTV and the RV techniques \citep{Mills17}.

K2-24 has two known transiting planets, which were observed by \Kepler during \ktwo operations \citep{Howell14}. \cite{Petigura16}, P16 hereafter, reported mass measurements based Keck/HIRES RVs spanning one observing season. While P16 predicted TTV amplitudes of several hours based on their proximity to the 2:1 mean-motion resonance, the 80~day \ktwo baseline was too short to observe deviations from linear ephemerides.

Here, we present an extended RV time series and additional transit-timing measurements from \spitzer (Section~\ref{sec:obs}). Our extended RV dataset enables tighter constraints on the planet masses and reveals a third candidate planet in the system (Section~\ref{sec:rv-keplerian}). In Section~\ref{sec:ttv}, we perform a joint TTV/RV analysis, which provides improved constraints on planet masses, eccentricities, and core/envelope fractions (Section~\ref{sec:constraints}). In Section~\ref{sec:dynamics}, we interpret the observed eccentricities in the context of system dynamics and formation scenarios, and we conclude in Section~\ref{sec:conclusions}.

\section{Observations}
\label{sec:obs}
\subsection{K2}
\label{sec:obs-ktwo}
\begin{figure*}
\centering
\includegraphics[width=0.9\textwidth]{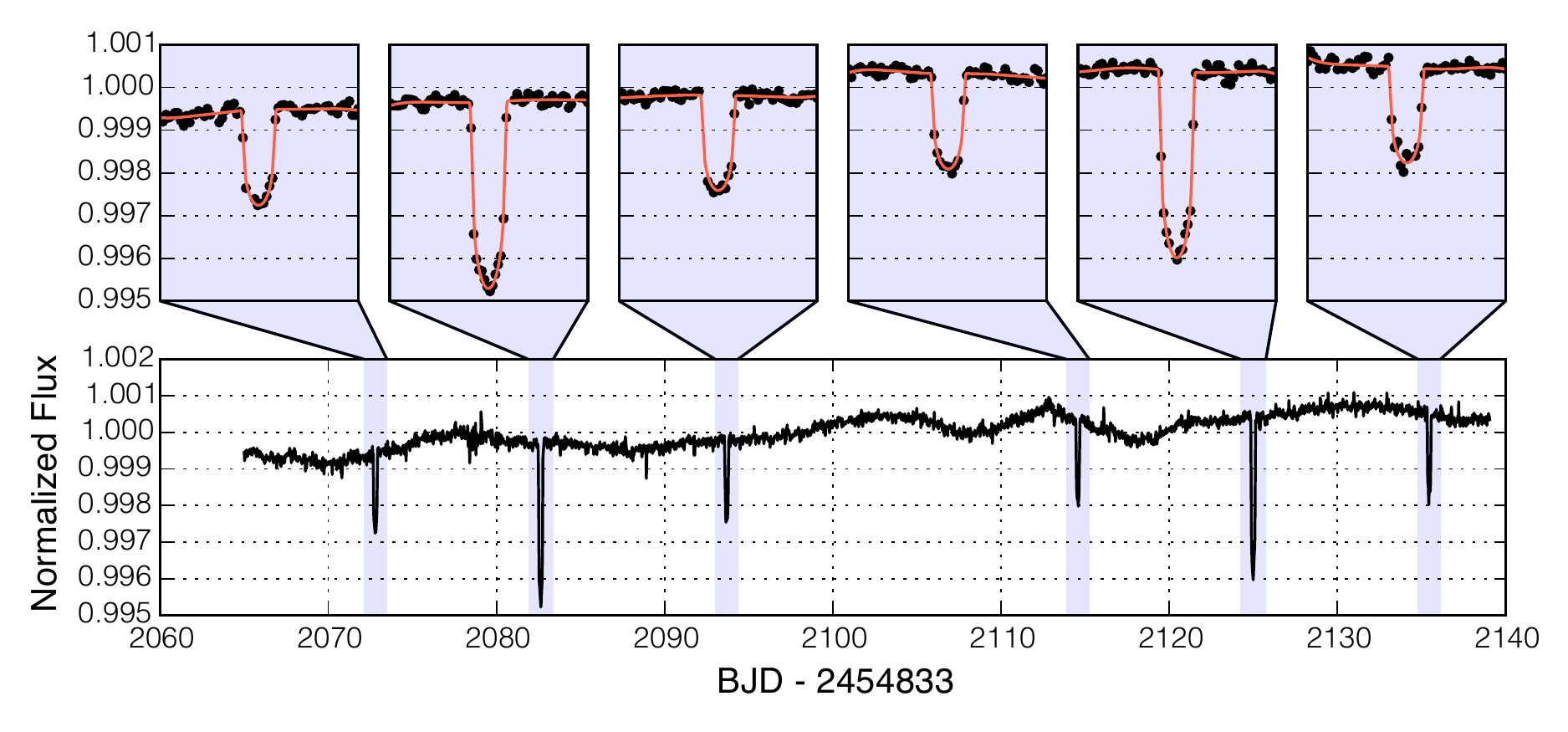}
\caption{Fits to the \ktwo photometry described in Section~\ref{sec:obs-ktwo}. The bottom panel shows the full \ktwo observing baseline from \cite{Petigura16}, and the insets show fits to individual transits.\label{fig:k2phot-transits}}
\end{figure*}

K2-24 was observed during campaign 2 of the \ktwo mission from 2014-08-23 to 2014-10-13. To extract transit times, we used the photometry published in P16 and fit individual transits. We multiplied our transit model by a third-order polynomial to account for the long timescale variability seen in the photometry. For each transit, we first adopted the best-fit transit parameters from P16, which assumed linear ephemerides. We then fit the transit allowing the time of conjunction \tc{} and the polynomial coefficients to vary. Figure~\ref{fig:k2phot-transits} shows the \ktwo photometry along with the best-fit transit models.

Care is required when assigning reasonable uncertainties to the measured transit times. \ktwo photometry contains correlated, non-Gaussian systematics that are mostly, but not entirely, removed during detrending.%
\footnote{For a more detailed discussion of \ktwo systematics, see \cite{Petigura18} and references therein.}
The derived transit times depend most sensitively on photometry collected during ingress or egress, which span one or two 30-minute long cadence measurements. Therefore, outliers have a significant effect on the derived transit times if they occur during ingress or egress. As an example, \cite{Benneke17} found that a single outlier that occurred during one of the transits of K2-18b resulted in a $\approx7\sigma$ error in the ephemeris reported in \cite{Montet15}. 

We estimated the \ktwo transit-timing errors errors via bootstrap resampling. For each transit, we created 1000 realizations by randomly shuffling the residuals to the best-fit light curve and adding the shuffled residuals to the best-fit model. We then fit these bootstrap realizations using the methods described above and derived \tc{} for each sample. We adopted the standard deviation of the resampled \tc{} as the uncertainty on \tc{}. The bootstrapped uncertainties were roughly twice as large as the formal uncertainties, which assumed white and Gaussian distributed noise. Our measured transit times are listed in Table~\ref{tab:transit-times}.

\begin{deluxetable}{llrrr}
\tablecaption{Transit Times\label{tab:transit-times}}
\tablehead{
  \colhead{Instrument} & 
  \colhead{Planet} & 
  \colhead{$i$} & 
  \colhead{$T_{c}$} & 
  \colhead{$\sigma(T_{c})$} \\
  \colhead{} & 
  \colhead{} & 
  \colhead{} & 
  \colhead{days} & 
  \colhead{days}
}
\startdata
      K2 &      b &       0 & 2072.7954 &  0.0011 \\
      K2 &      c &       0 & 2082.6248 &  0.0006 \\
      K2 &      b &       1 & 2093.6806 &  0.0013 \\
      K2 &      b &       2 & 2114.5654 &  0.0009 \\
      K2 &      c &       1 & 2124.9879 &  0.0006 \\
      K2 &      b &       3 & 2135.4505 &  0.0012 \\
 Spitzer &      b &      20 & 2490.6161 &  0.0011 \\
 Spitzer &      c &      10 & 2506.0002 &  0.0014 \\
 Spitzer &      c &      15 & 2717.5074 &  0.0015 \\
 Spitzer &      b &      31 & 2720.5049 &  0.0016 \\
\enddata
\tablecomments{Following a convention from the \Kepler mission, times are given in $\mathrm{BJD}_\mathrm{TBD} - 2454833$}
\end{deluxetable}

\subsection{Spitzer}
\label{sec:obs-spitzer}
P16 used analytic approximations developed by \cite{Lithwick12} to predict the expected TTVs of K2-24b and c. These approximations predicted anti-correlated sinusoidal TTVs having a ``super-period'' of roughly 4 years. Given the proximity of K2-24b and c to the 2:1 mean-motion resonance, P16 predicted large TTV amplitudes of several hours. However, the limited 80-day \ktwo baseline sampled only 5\% of the TTV super-period, too small a fraction for TTVs to accumulate to detectable levels. 

To cover a significant fraction of the expected TTV super-period, we used \spitzer to observe two additional transits of K2-24b on 2015-10-27 and 2016-06-13 and two additional transits of K2-24c on 2015-11-12 and 2016-06-10.%
\footnote{The 2015 observations were carried out under Director's Discretionary Time program 11184 (PI: M. Werner), while the 2016 observations were part of GO program 12107 (PI: E. Petigura).}
The combined \ktwo/\spitzer dataset includes transit observations at three well-separated epochs, which is sufficient to constrain the mean transit period as well as the amplitude and phase of the approximately sinusoidal TTV signal.

\begin{figure*}
\centering
\includegraphics[trim={0cm 6cm 14.5cm 0cm},width=0.75\textwidth]{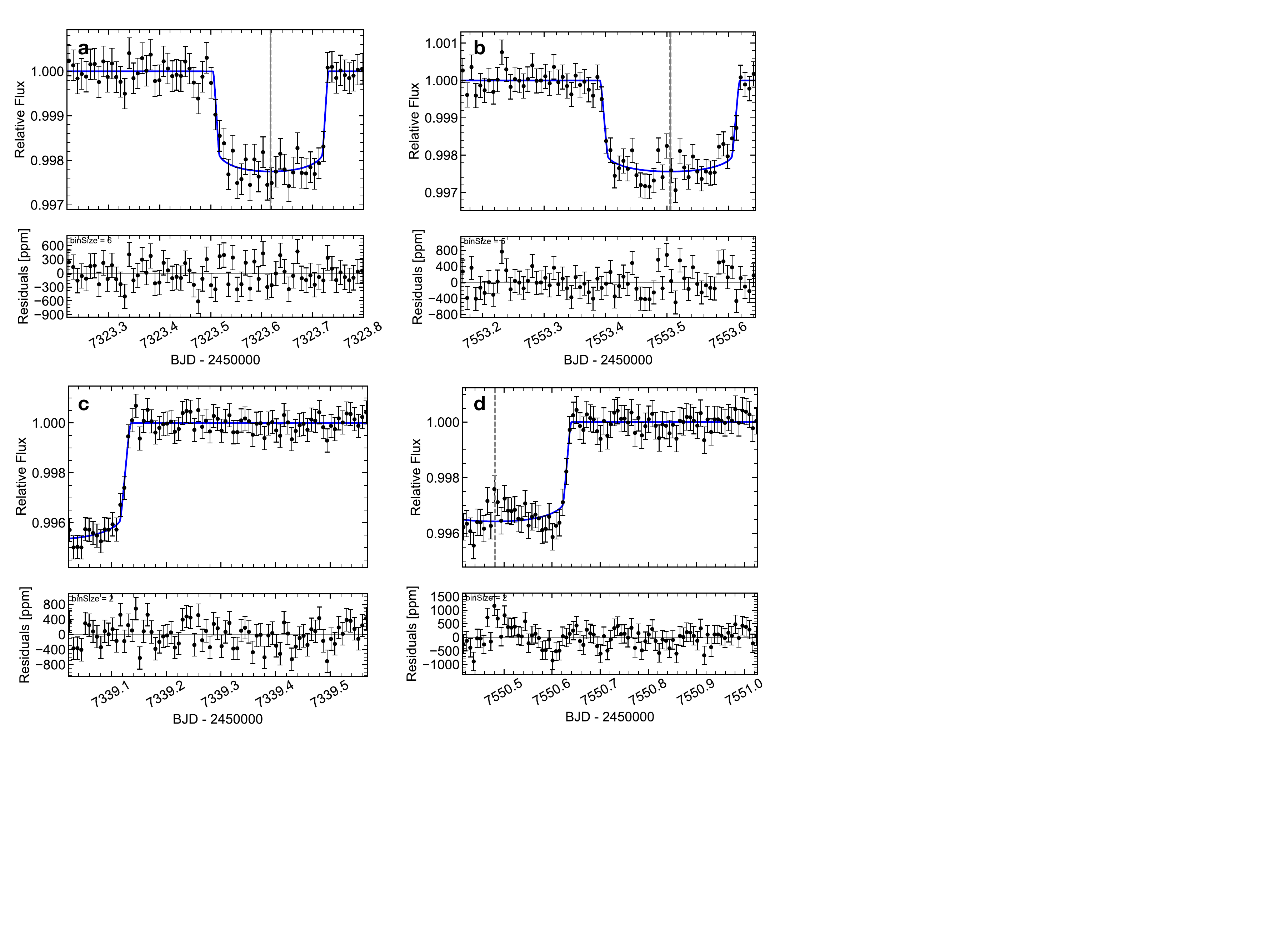}
\caption{Transits of K2-24b and c observed by \spitzer in the 4.5~\um IRAC channel. Panel (a) shows the first \spitzer observation of K2-24b transit number $i=20$, where $i=0$ corresponds to the first \ktwo transit. Points are the PLD-corrected photometry and the solid line is the most probable transit model. The transit is not centered in the \spitzer window due to TTVs of several hours. Panel (b): same as (a) but for the second \spitzer observation of K2-24b ($i=31$).  Panel (c): same as (a) but for the first \spitzer observation of K2-24c ($i=10$). Panel (d): same as (a) but for the second \spitzer observation of K2-24c ($i=15$).\label{fig:lightcurve-spitzer}}
\end{figure*}

When planning our 2015 \spitzer observations, we centered our observing sequence using the best-fit transit times of K2-24b and c based on the \ktwo data alone. To account for the substantial uncertainty due to TTVs, we observed K2-24b and c for 14 hours each. As shown in Figure~\ref{fig:lightcurve-spitzer}, we observed a complete transit of K2-24b and a partial transit of K2-24c. We centered our 2016 \spitzer observations the best-fit linear ephemeris that incorporated the \ktwo and 2015 \spitzer observations, and we observed K2-24b and c for 12 and 16 hours, respectively. Again, we observed a complete transit of K2-24b and a partial transit of K2-24c. In hindsight, after collecting the 2015 \spitzer transits we should have performed a preliminary TTV model using plausible masses and eccentricities in order to better center our 2016 \spitzer observations.

Following common practice, we included a 30-minute pre-observation sequence to mitigate the initial instrument drift in the science observations resulting from telescope temperature changes after slewing from the preceding target \citep{Grillmair12}. To enhance the accuracy in positioning K2-24 on the IRAC detector, observations were taken in peak-up mode using the Pointing Calibration and Reference Sensor (PCRS) as a positional reference. We chose Spitzer/IRAC Channel 2 (4.5~$\mu$m) over Channel 1 (3.6~$\mu$m) because the instrumental systematics due to intra-pixel sensitivity variations are smaller \citep{Ingalls12}. Our exposure times were set to 2 seconds to optimize the integration efficiency while remaining in the linear regime of the IRAC detector.

Following  \cite{Benneke17}, we extracted multiple photometric light curves for each \spitzer dataset using a wide range of fixed and variable aperture sizes. The purpose of extracting and comparing multiple photometric light curves is to choose the aperture that provides the lowest residual scatter and red noise. We normalized the light curve by the median value and binned the data to a 60-second cadence. We found that this moderate binning did not affect the information content of the photometry, but provided more signal per data point allowing an improved correction of the systematics.

Raw aperture photometry from \spitzer contains large systematics due to the motion of the target star across the IRAC detector with percent-level intra-pixel sensitivity variations. To extract reliable transit times, we adopted the standard practice of modeling the \spitzer systematics and transit profile simultaneously. We used the pixel-level decorrelation (PLD) algorithm, first proposed by \cite{Deming15}, with modifications described in \cite{Benneke17}. 

In our model, the following transit parameters were allowed to vary: transit midpoint \tc{}, planet-to-star radius ratio \Rp/\Rstar, and impact parameter $b$. In addition, we parameterized the systematics in the \spitzer model using nine PLD coefficients, a white noise component, and two coefficients describing a polynomial trend of flux with time. Ideally, we would have allowed the transit duration \T{14} to vary in our fits. However, because our \spitzer transit observations of K2-24c missed ingress, they could not meaningfully constrain \T{14}. For both K2-24b and c, we fixed \T{14} to the value measured by P16 from \ktwo photometry. We explored the likelihood surface using Markov Chain Monte Carlo (MCMC). The maximum likelihood fits to the \spitzer photometry are shown in Figure~\ref{fig:lightcurve-spitzer}, and the associated transit times are listed in Table~\ref{tab:transit-times}.

\subsection{Keck/HIRES Spectroscopy}
\label{sec:obs-hires}
We obtained \stat{rv-n} spectra of K2-24 using the High Resolution Echelle Spectrometer (HIRES; \citealt{Vogt94}) on the 10m Keck-I telescope between \stat{rv-start} and  \stat{rv-stop}. We collected spectra through an iodine cell mounted directly in front of the spectrometer slit. The iodine cell imprints a dense forest of absorption lines which serve as a wavelength reference. We used an exposure meter to achieve a consistent signal-to-noise level of 110 per reduced pixel on blaze near 550 nm. We also obtained a ``template'' spectrum without iodine. The first 32 of these spectroscopic observations are described in P16.

RVs were determined using standard procedures of the California Planet Search \citep{Howard10b} including forward modeling of the stellar and iodine spectra convolved with the instrumental response (\citealt{Marcy92,Valenti95}). The measurement uncertainty of each RV point is derived from the uncertainty on the mean RV of the $\sim$700 spectral chunks used in the RV pipeline and ranges from \stat{rv-errvel-min} to \stat{rv-errvel-max} \ms. Table~\ref{tab:rv} lists the RVs and uncertainties.

\begin{deluxetable}{RRRRRr}
\tablecaption{Radial Velocities\label{tab:rv}}
\tablecolumns{5}
\tablewidth{-0pt}
\tabletypesize{\footnotesize}
\tablehead{
  \colhead{Time} & 
  \colhead{RV} & 
  \colhead{$\sigma$(RV)} &
  \colhead{$\mathrm{S}_\mathrm{HK}$} \\ 
  \colhead{days} & 
  \colhead{\ms} & 
  \colhead{\ms} & 
  \colhead{}  
}
\startdata
2364.819580 & 0.85 & 1.68 & 0.132 \\
2364.825101 & 1.72 & 1.52 & 0.130 \\
2364.830703 & 9.99 & 1.59 & 0.132 \\
2366.827579 & -3.90 & 1.62 & 0.128 \\
2367.852646 & 5.50 & 1.65 & 0.130 \\
2373.888150 & -3.77 & 1.78 & 0.094 \\
2374.852412 & -5.65 & 1.97 & 0.113 \\
2376.863820 & -6.09 & 1.79 & 0.131 \\
2377.866073 & -2.40 & 1.76 & 0.131 \\
2378.834011 & -1.33 & 1.60 & 0.131 \\
\enddata
\tablecomments{Radial velocities and uncertainties for K2-24. Times are given in $\mathrm{BJD}_\mathrm{TBD} - 2454833$.  We also provide the Mount Wilson $\mathrm{S}_\mathrm{HK}$ activity index \citep{Vaughan78}, which is measured to 1\% precision. Table \ref{tab:rv} is published in its entirety in machine-readable format. A portion is shown here for guidance regarding its form and content.}
\end{deluxetable}

\section{RV analysis}
\label{sec:rv-keplerian}
\begin{figure}
\centering
\includegraphics[trim=0.1cm 0.0cm 0.1cm 0cm, clip,width=0.48\textwidth]{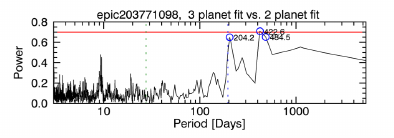}
\label{fig:rv-keplerian}
\caption{Searches for Keplerian signatures in the HIRES RV time series of K2-24 after removing contributions from K2-24b and c using a Two-Dimensional Keplerian Lomb-Scargle periodogram. We observe a peak at $P = 420$~days and its first harmonic.\label{fig:rv-keplerian-search}}
\end{figure}

Here we present our Keplerian analysis of the K2-24 RVs.  The RVs exhibited $\approx$10~\ms peak-to-trough variability that was not associated with the known ephemerides of K2-24b or c, which motivated searches for additional non-transiting planets. Figure~\ref{fig:rv-keplerian-search} shows a Keplerian search using a modified version of the Two-Dimensional Keplerian Lomb-Scargle (2DKLS) periodogram \citep{Otoole09,Howard16}. When we measured the change in $\chi^{2}$ (periodogram power) between a three-planet fit and a two-planet fit, we found a peak at $P$ = 420~days, with an empirical false alarm probability (eFAP) of 0.8\%. While the eFAP was formally below the standard criterion of eFAP < 1\% for Doppler confirmation, a complete confirmation of this candidate would have required additional vetting such as an assessment of RV/activity correlations, which is beyond the scope of this work. We included this candidate our subsequent orbit fitting because it improved the quality of the RV fits to K2-24b and c.

We analyzed the RV timeseries using the publicly available RV modeling package \radvel \citep{Fulton18a}. \radvel facilitates maximum a posteriori (MAP) model fitting and parameter estimation via MCMC. A Keplerian RV signal may be described by the orbital period $P$, time of inferior conjunction \tc{}, eccentricity $e$, longitude of periastron $\omega$ and Doppler semi-amplitude $K$, i.e. $\{P, \tc{}, e, \omega, K \}$. In our fitting and MCMC analysis, we adopted the following parameterization: $\{ \ln P, \tc{}, \secosw{}, \sesinw{}, \K{} \}$. Our parameterization of $e$ and $\omega$ enforces a uniform prior on eccentricity and prevents a Lucy-Sweeney bias toward non-zero eccentricities \citep{Eastman13,Fulton18a}. Our preferred model consists of three Keplerians with eccentricities fixed to zero. We fixed the $P$ and $T_{c}$ of K2-24b and c to the P16 values. To aid convergence, we imposed a loose Gaussian prior on $\ln P_{d}$ of $\mathcal{N}(\ln(440),1)$. Figure \ref{fig:rv-keplerian} shows the MAP model.

Models with more free parameters will naturally lead to higher likelihoods at the expense of additional model complexity. To compare the quality of models of different complexity we used the Bayesian Information Criterion (BIC; \citealt{Schwartz78}). Models with smaller BIC are preferred. For the circular, three planet model, BIC = \hand{fit-ccc-bic}. Models where candidate d is allowed to have a non-zero eccentricity were not favored by the BIC = \hand{fit-cce-bic}. Models with only two planets on circular orbits were also disfavored by the BIC = \hand{fit-cc-bic}.

To derive uncertainties on the model parameters, we used \radvel to sample the posterior probability via MCMC. \radvel automatically checks for convergence using the Gelman-Rubin statistic \citep{Gelman92}. For K2-24b and c, our RV only analysis yields masses of \fit{ccc-mpsini1_fmt}~\Me and \fit{ccc-mpsini2_fmt}~\Me, respectively. We compare these masses to those determined by the joint TTV/RV analysis in Section~\ref{sec:constraints}. If candidate d is a bonafide planet, it has a mass of \fit{ccc-mpsini3_fmt}~\Me and orbits at a distance of \fit{ccc-a3_fmt}~AU. However, we do not treat candidate d in our subsequent analysis or discussion, because we have not performed a thorough confirmation and because it is decoupled dynamically from the inner two planets.

Even though the model with all three eccentricities set to zero was preferred in a BIC sense, we performed an analogous MCMC exploration with eccentric orbits to asses the extent to which the RVs alone constrain eccentricities. The RV dataset only ruled out high eccentricity orbits, with upper limits of \ecc{1} < \fit{eec-e1_p90} and \ecc{2} < \fit{eec-e2_p90} at 90\% confidence.

\begin{figure*}
\centering
\includegraphics[width=0.9\textwidth]{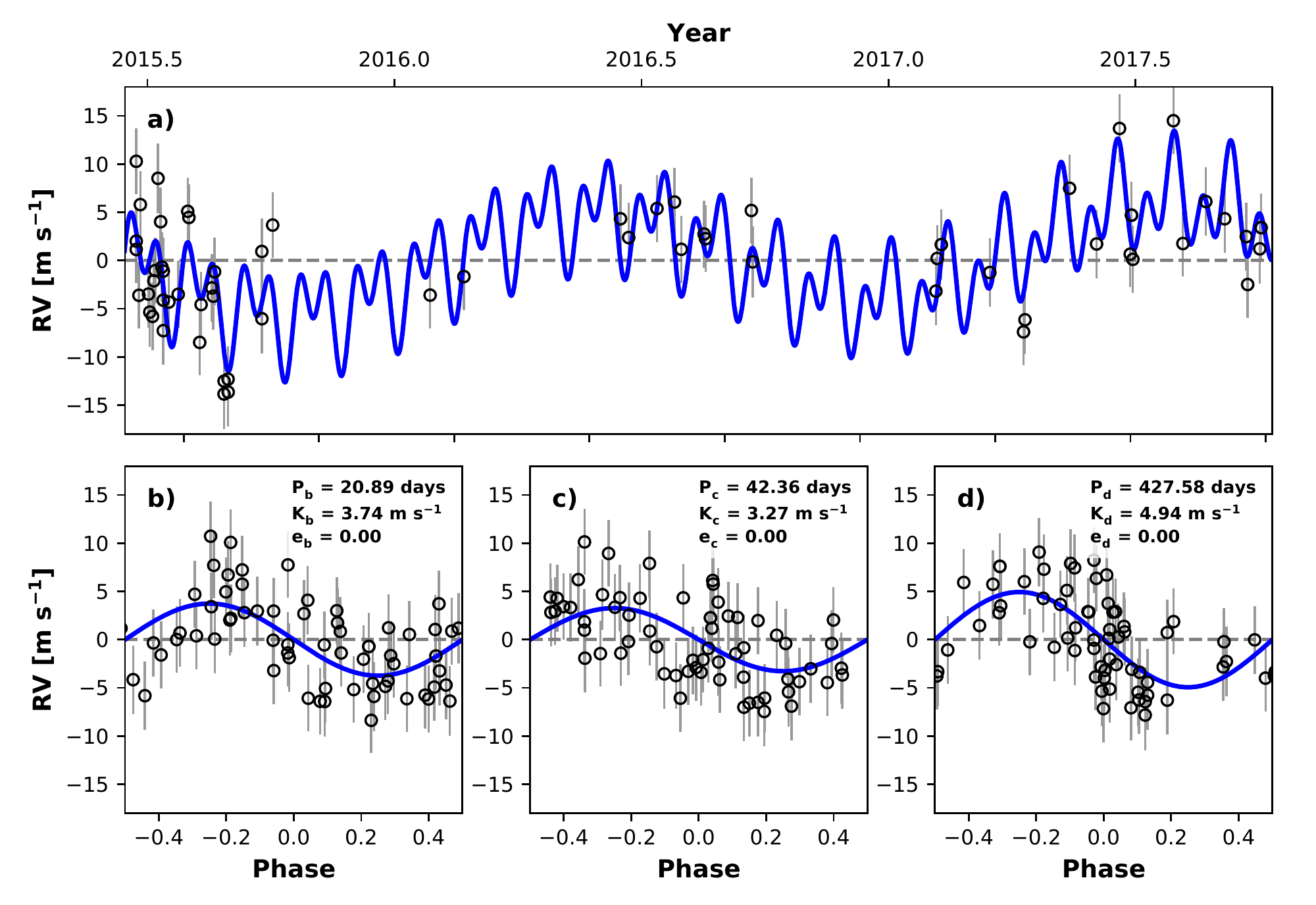}
\label{fig:rv-keplerian}
\caption{The three-Keplerian fit to the K2-24 radial velocities (RVs), assuming circular orbits described in Section~\ref{sec:rv-keplerian}. Panel (a): Points show RVs from HIRES and the line shows the most probable Keplerian model. Panel (b) shows the phase-folded RVs and the most probable Keplerian model for K2-24b with contributions from other Keplerians removed. Panel (c)--(d), same as (b), but for K2-24c and candidate d.\label{fig:rv-keplerian}}
\end{figure*}

\section{Joint TTV/RV Analysis}
\label{sec:ttv}
As expected, the \spitzer observations revealed TTVs of several hours. In this section, we present an analysis of the transit times from \ktwo and \spitzer, folding in the constraints from RVs described in the previous section. 

\cite{Lithwick12}, L12 hereafter, developed an analytical model for the TTVs that occur when two planets are near first order mean-motion resonance (i.e., $P_2$:$P_1$ $\approx$ $j$:$j-1$, where $j = 2, 3, \dots$). For a complete exposition of this formalism, see L12. Here, we provide a brief summary, in order to illustrate the type of constraints that the TTVs provide.

For planets near, but not in, first order mean-motion resonance L12 showed that their transit times, \tc{i}, are described by a sinusoidal perturbation about a mean period, $P$:
\begin{equation}
\label{eqn:lithwick}
\tc{i} = T_{c,0} + P i +\mathrm{Re}(V) \sin \lambda_{j} +\mathrm{Im}(V) \cos \lambda_j.
\end{equation}
Here, $i$ is an integer index that labels the transit epoch, $T_{c,0}$ is the time of the first transit ($i = 0$), and $V$ is the complex TTV amplitude. The longitude of conjunctions
$\lambda_j$, is an angle that advances linearly with time and is given by
\begin{eqnarray}
\lambda_j & = & j \lambda_2 - (j - 1) \lambda_1, \\ 
\lambda_1   & = & \frac{2 \pi}{P_1}\left(t - \tc{1} \right), \\
\lambda_2   & = & \frac{2 \pi}{P_2}\left(t - \tc{2} \right). 
\end{eqnarray}
The time it takes $\lambda_j$ to advance by $2\pi$ is known as the super-period $P_j$, which is given by
\begin{eqnarray}
P_j  & \equiv &  \frac{P_{2}}{j |\Delta|}, \\
\Delta & \equiv &  \frac{P_{2}}{P_{1}}\frac{j - 1}{j} - 1.
\end{eqnarray}
For the K2-24bc pair, $\Delta = 0.013$ and $P_j = 1595$~days. The complex TTV amplitudes are given by
\begin{eqnarray}
V_1 & = & P_1 \frac{\mu_2}{\pi j^{2/3} (j-1)^{1/3}\Delta} \left(-f - \frac{3}{2} \frac{\Zfreestar}{\Delta}\right) \label{eqn:V1}\\
V_2 & = & P_2 \frac{\mu_1}{\pi j \Delta}\left(-g + \frac{3}{2} \frac{\Zfreestar}{\Delta} \right)\label{eqn:V2},
\end{eqnarray}
respectively, where $\mu$ is the planet-star mass ratio and $f$ and $g$ are order unity scalar coefficients which depend $j$ and $\Delta$ and are given in L12. For the K2-24bc, $f = -1.16$ and $g = 0.38$.  \Zfreestar is the complex conjugate of the following linear combination of the planets complex eccentricities:
\begin{equation}
\label{eqn:zfree}
\Zfree = f \zfree{1} + g\zfree{2},
\end{equation}
where 
\begin{equation}
z = e \cos \varpi + i e \sin \varpi.
\end{equation}
Our full TTV model contains the following free parameters: $\{ P_1, \tc{1}, \mu_1, P_2, \tc{2}, \mu_2, \mathrm{Re}(\Zfree), \mathrm{Im}(\Zfree) \}$.

We incorporated Gaussian priors of $\mu_1 = \stat{lithwick-prior-mu1} \pm \stat{lithwick-prior-mu1_err}$~ppm and $\mu_2 = \stat{lithwick-prior-mu2} \pm \stat{lithwick-prior-mu2_err}$~ppm based on our RV analysis in Section~\ref{sec:rv-keplerian}. We confirmed that Gaussian priors were appropriate by checking that the RV-only constraints on $\mu_1$ and $\mu_2$ are well-described by normal distributions, with negligible covariance (Pearson $r$ = 0.09).  

We explored the range of plausible planet masses and orbits given the measured transit times using the Affine-Invariant MCMC sampler of \cite{Goodman10}. We found that employing parallel tempering dramatically reduced the number of iterations needed for convergence \citep{Earl05}. We let 16 walkers evolve for 50,000 iterations at five different temperatures, discarding the first 10,000 iterations as burn in. We verified that the chains were well-mixed by computing the autocorrelation length scale $\tau$ for each chain at each temperature and confirming that $\tau$ is much smaller than the number of iterations.

\begin{figure*}
\centering
\includegraphics[width=0.8\textwidth]{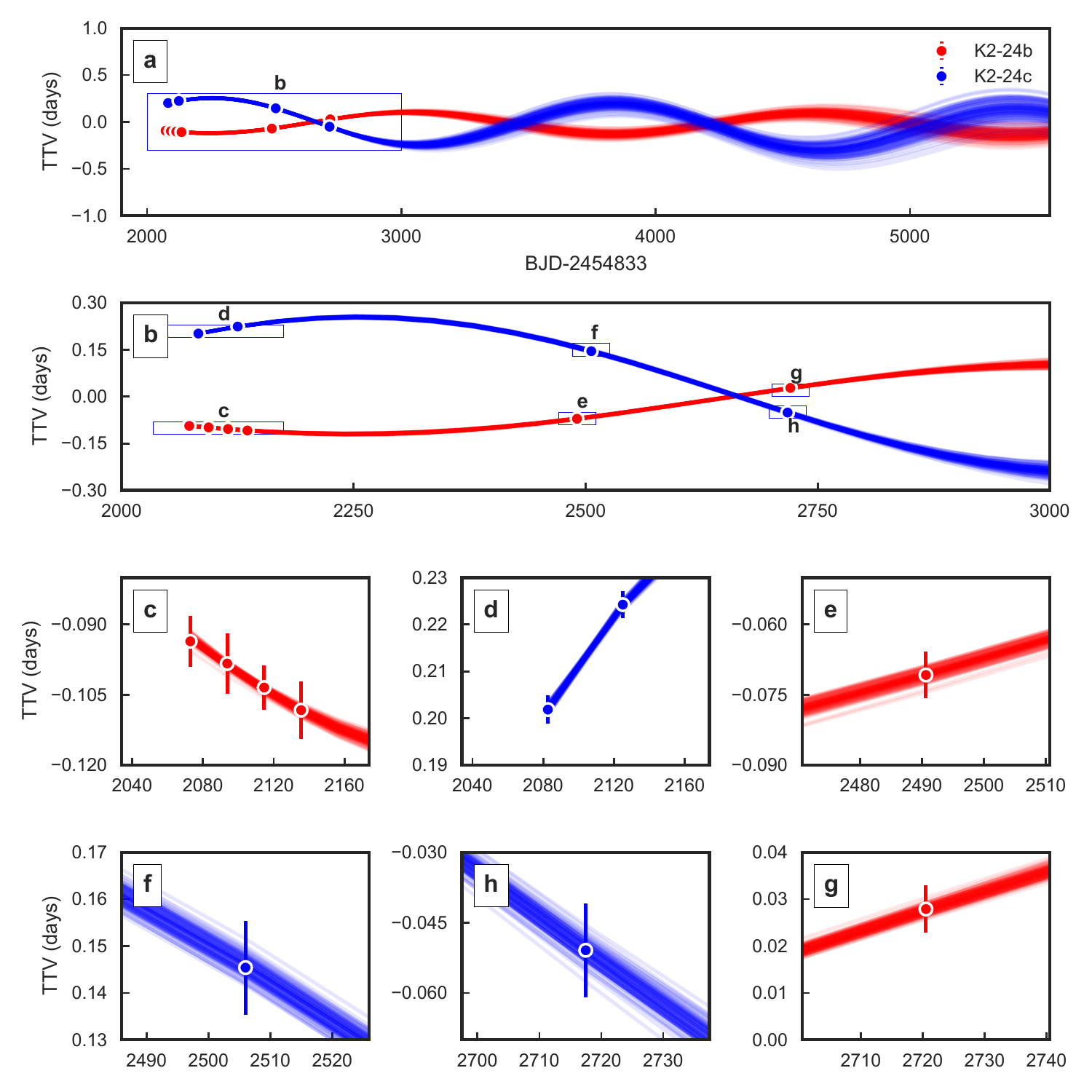}
\caption{Points show transit-timing variations (TTVs) of K2-24b and c with respect to linear ephemerides (Section~\ref{sec:obs}). Lines show TTV models based on 100 draws from the MCMC samples, explained in Section~\ref{sec:ttv}. Panel (a) shows 10 years of predicted TTVs. Panel (b) same as (a), but showing the TTV behavior over the baseline of \ktwo and \spitzer observations. Panels (c)--(f) highlight the model fits around individual transit epochs. The errorbars have been enlarged by a factor of 5 for legibility.\label{fig:lithwick-samples}}
\end{figure*}

In Figure~\ref{fig:lithwick-samples}, we display the measured and modeled transit times with respect to an adopted reference linear ephemeris. 
%
%
The models sampled from the posterior are a good fit to the observed transit times and gradually diverge from one another after the last \spitzer measurement. To facilitate future observations of K2-24b and c, we include the predicted transit times and uncertainties through 2025 in the Appendix.

Figure~\ref{fig:lithwick-corner4} shows the two-parameter joint posterior distributions. Note the strong covariance between $\mu_1$ and $\mu_2$. As expected, the TTVs enabled a tight constraint on the planet mass ratio of $\mr$ = \lithwick{mr2_fmt}. As a point of comparison, the RV-only fits constrained the mass ratio to $\mr = 1.10_{-0.26}^{+0.34}$, which is consistent at the $1\sigma$ level.

Note also the strong covariance between $\mu$--\Zfree. The priors on $\mu_{1}$ and $\mu_{2}$ help to break the $\mu$--\Zfree degeneracy, and we detect significant non-zero real imaginary components of \Zfree. While \Zfree only constrains linear combinations of the eccentricities, we could infer that (1) at least one of the planets has a non-zero eccentricity and (2) the eccentricities are likely $|\Zfree | \sim 0.08$. Recall that the RV analysis in Section~\ref{sec:rv-keplerian} only provided upper limits of $e_{1} < \fit{eec-e1_p90}$ and $e_{2} < \fit{eec-e2_p90}$. Because the TTVs constrain only linear combinations of the $e_1$ and $e_2$, we cannot rule out high eccentricity solutions. However, as we discuss in Section~\ref{sec:constraints}, these solutions are unlikely given the low eccentricities typically observed in compact \Kepler multi-planet systems.

\begin{figure}
\centering
\includegraphics[width=0.5\textwidth]{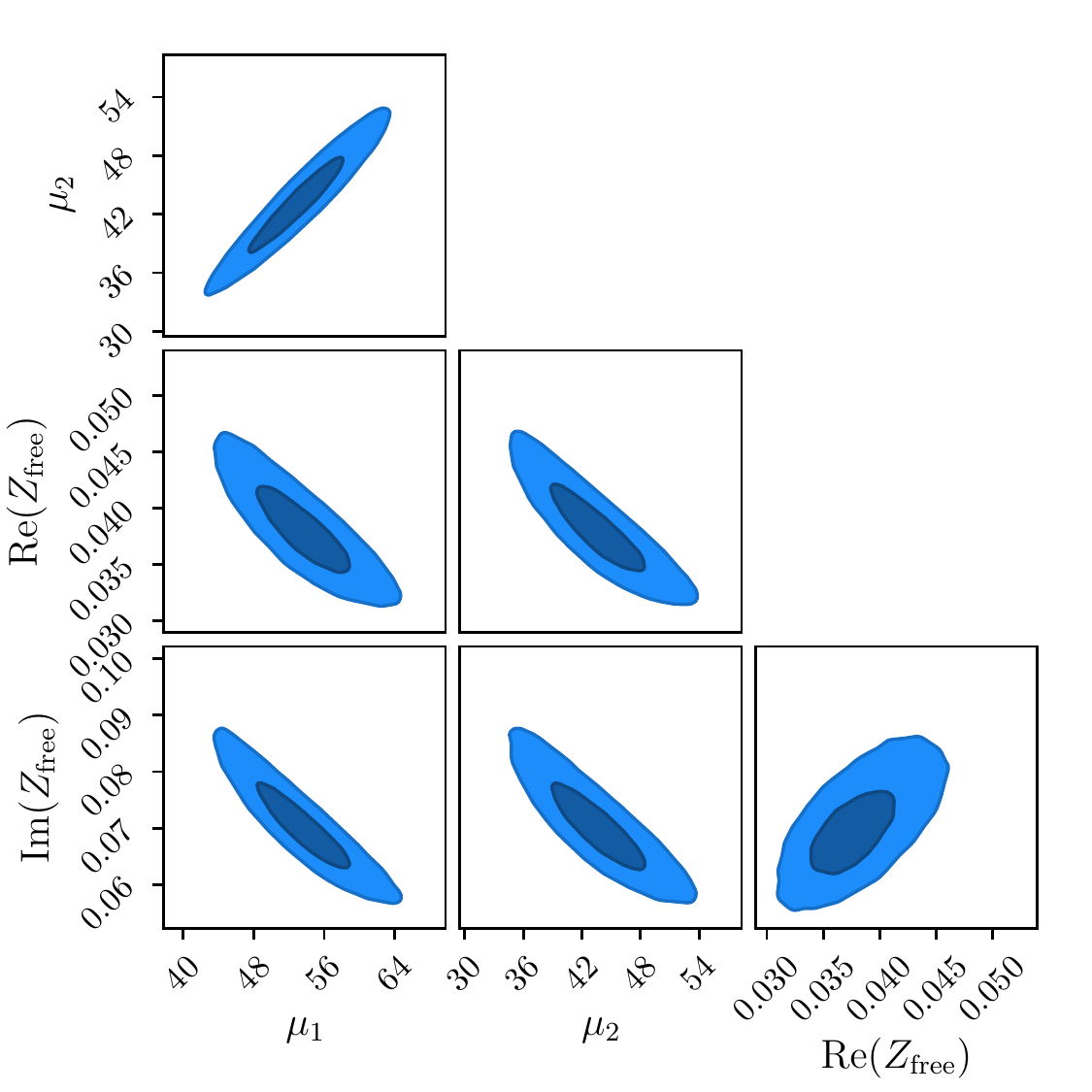}
\caption{Constraints on $\mu_1$, $\mu_2$, $\ReZfree$, and $\ImZfree$ given our TTV/RV analysis (Section~\ref{sec:ttv}). Contours show $1\sigma$ and $2\sigma$ levels. This modeling produced tight constraints on  $\mu_2/ \mu_1  = \mr$ and on \Zfree.\label{fig:lithwick-corner4}}
\end{figure}

{\renewcommand{\arraystretch}{0.9}
\begin{table}
\begin{center}
\caption{K2-24 System Parameters}
\begin{tabular}{lrr}
\hline
\hline
Parameter              & Value   & Notes \\
\hline
\multicolumn{2}{l}{{\bf Stellar Parameters}} \\
\teff (K)               & \stellar{br-teff}      & A \\
\logg (dex)             & \stellar{br-logg}      & A \\
\fe (dex)               & \stellar{br-fe}        & A \\
$K$ (mag)               & $9.18 \pm 0.02 $       & B \\
$\pi_{\star}$ (mas)     & $5.84 \pm 0.05 $       & C \\
\Mstar (\Msun)          & \stellar{br-mass}      & D \\
\Rstar (\Rsun)          & \stellar{br-radius}    & D \\
\multicolumn{2}{l}{{\bf Model Parameters}} \\
$P_1$ (days)              & \lithwick{per1_fmt}    & E \\
$\tc{1}$ (BJD$-$2454833)  & \lithwick{tc1_fmt}     & E \\
$\mu_1$ (ppm)             & \lithwick{muppm1_fmt}  & E \\
$P_2$ (days)              & \lithwick{per2_fmt}    & E \\
$\tc{2}$ (BJD$-$2454833)  & \lithwick{tc2_fmt}     & E \\
$\mu_2$ (ppm)             & \lithwick{muppm2_fmt}  & E \\
\ReZfree                  & \lithwick{rezfree_fmt} & E \\
\ImZfree                  & \lithwick{imzfree_fmt} & E \\
\multicolumn{2}{l}{{\bf Derived Parameters}} \\
$|\Zfree|$                & \lithwick{zmag_fmt}    & F \\
$\mr$                     & \lithwick{mr2_fmt}     & F \\
$M_{p,1}$ (\Me)           & \lithwick{masse1_fmt}  & F \\
$M_{p,2}$ (\Me)           & \lithwick{masse2_fmt}  & F \\
$R_{p,1}$ (\Re)           & \lithwick{prad1_fmt}   & F \\
$R_{p,2}$ (\Re)           & \lithwick{prad2_fmt}   & F \\
$\rho_1$ (g cm$^{-3}$)    & \lithwick{rho1_fmt}    & F \\
$\rho_2$ (g cm$^{-3}$)    & \lithwick{rho2_fmt}    & F \\
$e_1$                     & \lithwick{e1_fmt}      & G \\
$e_2$                     & < \lithwick{e2_p90} (90\% conf.) & G \\
\hline
\\[-6ex]
\end{tabular}
\end{center}
\tablecomments{A: \cite{Brewer16}. B: 2MASS \citep{Skrutskie06}. C: {\em Gaia} DR2 \citep{Gaia18}. D: Derived from A, B, and C using the methodology described in \cite{Fulton18b}. E: See Section~\ref{sec:ttv}. F: Derived from the posterior samples of D and E. G: Same as F, but with the eccentricity prior described in Section~\ref{sec:constraints}.}
\label{tab:lithwick}
\end{table}
}

\section{TTV/RV Synergies}
\label{sec:constraints}
In the previous section, we presented a joint TTV/RV analysis of the K2-24 system. Here, we provide an updated assessment of planet properties based on our combined TTV/RV analysis in Section~\ref{sec:ttv} and compare them to those presented in P16, which only included RVs. Orbital eccentricities are substantially improved over P16, and we also improve planet mass precision and constraints on core/envelope structures.

\subsection{Planet Mass}
P16 measured masses of K2-24b and c based on one season of RV measurements and found $M_{p,1}$ = \hand{masse1-p16_fmt}~\Me and $M_{p,2}$ =  \hand{masse2-p16_fmt}~\Me, respectively. Our analysis here yields masses of $M_{p,1}$ = \lithwick{masse1_fmt}~\Me and $M_{p,2}$ =  \lithwick{masse2_fmt}~\Me, respectively. The mass measurements from the two papers are consistent to within $2\sigma$, but our new masses have higher precision. The improved mass constraints are due to two factors: (1) more RV measurements with better phase coverage and (2) the strong constraint on \mr from the TTVs. Our TTV/RV analysis demonstrates that K2-24c is 20\% less massive than K2-24b, despite being 40\% larger.

\subsection{Core/Envelope Structure}
\cite{Petigura17a} examined the distribution of core masses $\mcore$  and envelope masses $\menv$ in a sample of 20 sub-Saturns (\Rp = 4--8~\Re), which included K2-24b and c. Planets in this size range are well-approximated by a two-component model consisting of a high density core and a thick envelope of near solar composition H/He. \cite{Lopez14} constructed a grid of model planets having different \mcore and \menv and computed their radii given different levels of stellar irradiation. For each planet in the sample, \cite{Petigura17a} used the \cite{Lopez14} grid to derive the range of \mcore and \menv consistent with the observed planet mass and radii. 

For K2-24b and c, \cite{Petigura17a} derived envelope fractions of $f_\mathrm{env,b}$ = \hand{fenv1-p17_fmt}\% and $f_\mathrm{env,c}$ = \hand{fenv2-p17_fmt}\%. We repeated this analysis using the updated planet masses and radii and found $f_\mathrm{env,b}$ = \lithwick{fenv1_fmt}\% and $f_\mathrm{env,c}$ = \lithwick{fenv2_fmt}\%. Our new values are consistent with \cite{Petigura17a}, but with smaller formal uncertainties. This stems mainly from the improved stellar radius (see Table~\ref{tab:lithwick}) and from the fact that, in the sub-Saturn size range, radius alone is a good proxy for envelope fraction \citep{Lopez14}.

One challenge in explaining the formation of K2-24c is to determine how the planet acquired such a large envelope, while avoiding runaway accretion. As a point of reference, in the canonical core accretion models of \cite{Pollack96}, Saturn forms first as a $\approx$12~\Me core that accretes H/He from the protoplanetary disk. At the crossover mass (i.e. when $\menv \approx \mcore$ or when $\fenv \approx 50\%$), runaway accretion begins and Saturn quickly grows to its final mass.

One way to resolve the $\fenv \approx 50\%$ problem is to imagine that the disk dissipated right as K2-24c approached the runaway phase. While impossible to rule out, this scenario requires special timing of planet formation and is thus a priori unlikely. More likely, the inferred structure of K2-24c points to an incomplete understanding of core-nucleated accretion and motivates further theoretical explanations of planet conglomeration in the sub-Saturn mass regime.

\subsection{Eccentricity}
By combining TTVs and RVs, we achieved significantly tighter constraints on eccentricity than those from either technique alone. The full RV dataset only provided weak upper limits on the planet eccentricities of $\ecc{1}$ < \fit{eec-e1_p90} and $\ecc{2}$ < \fit{eec-e2_p90}. The TTVs, in contrast, constrained $\mu_1 \Zfree$ and  $\mu_2 \Zfree$ (Equations~\ref{eqn:V1}--\ref{eqn:V2}). Because RVs constrain planet mass directly, they break some of the $\mu$--$\Zfree$ degeneracy inherent to a TTV-only analysis. 

\begin{figure}
\centering
\includegraphics[width=0.49\textwidth]{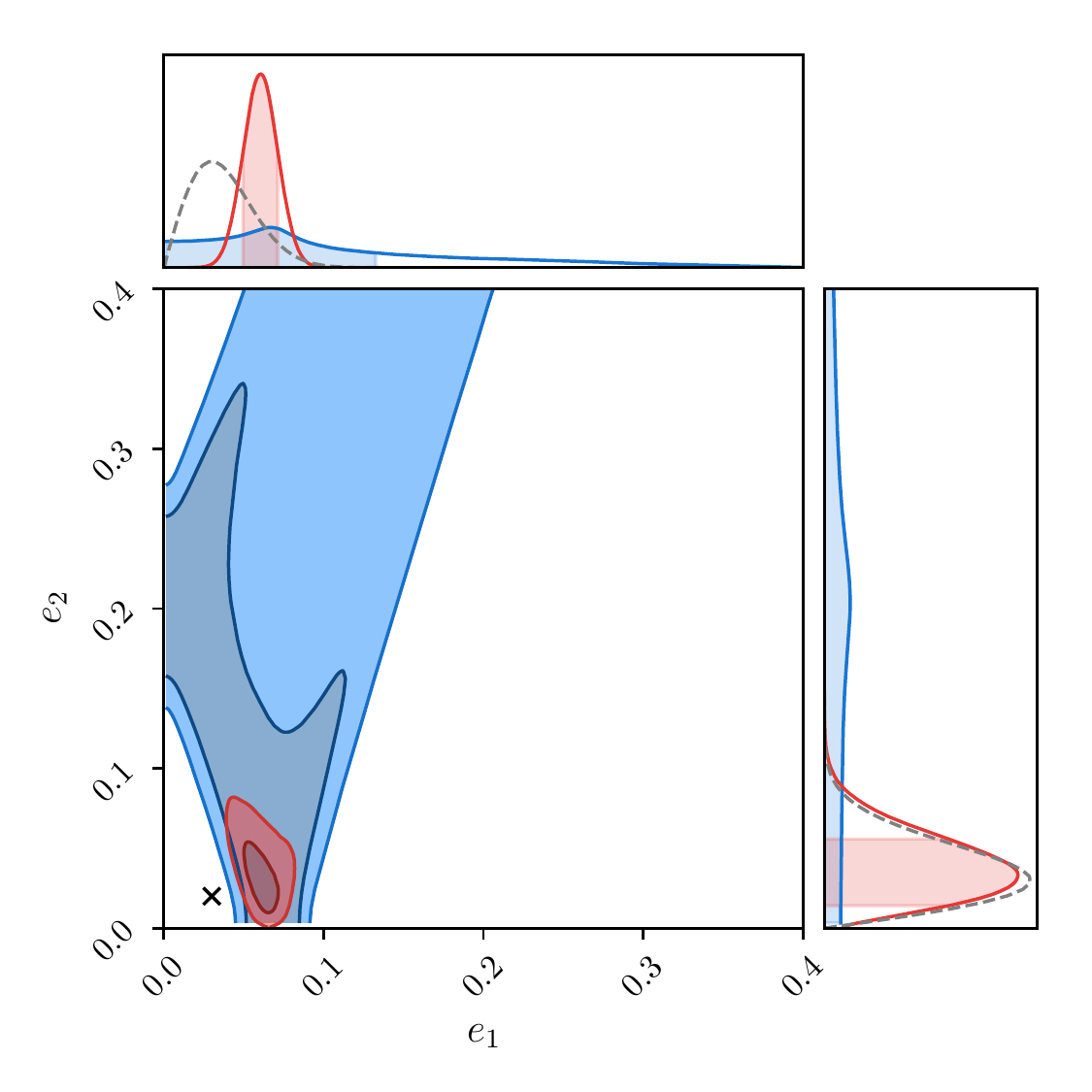}
\caption{The blue contours show the joint constraints on $e_1$ and $e_2$ from the TTV/RV analysis described in Section~\ref{sec:ttv}. Because the TTVs only constrain linear combinations of the eccentricities, a large range of $e_1$ and $e_2$ is consistent with the data. Note, however, $e_1$ and $e_2$ may cannot both be zero. The red contours incorporate a Rayleigh prior on eccentricities with $\langle e\rangle = 0.03$, which is shown as gray dotted lines in the 1D distributions. This prior is motivated in Section~\ref{sec:constraints}. Under this prior, solutions where $e_1 \sim 0.0$ are disfavored because they imply that $e_2 \sim 0.2$. The `x' marks $(e_1,e_2) = (0.02,0.03)$, which is expected if the system had experienced divergent migration through resonance (Section~\ref{sec:origin-eccentricity}).}
\label{fig:e1-e2}
\end{figure}

Our TTV/RV model provided the following constraints on $\ReZfree$ and $\ImZfree$:
\begin{eqnarray}
\ReZfree = & f e_1 \cos \varpi_1 + g e_2 \cos \varpi_2 & = \lithwick{rezfree}^{+\lithwick{rezfree_err1}}_{\lithwick{rezfree_err2}}\,\nonumber \\ 
\ImZfree = & f e_1 \sin \varpi_1 + g e_2 \sin \varpi_2 & = \lithwick{imzfree}^{+\lithwick{imzfree_err1}}_{\lithwick{imzfree_err2}}.\,\nonumber 
\end{eqnarray}
These constraints amount to lines in the \mbox{$e_1 \cos \varpi_1$-$e_2 \cos \varpi_2$} and \mbox{$e_1 \sin \varpi_1$-$e_2 \sin \varpi_2$} planes with slopes determined by $f$ and $g$. Because TTVs only constrain linear combinations of $e_1$ and $e_2$ there are still significant $e_1$-$e_2$ degeneracies, even after folding in the RV constraints. Figure~\ref{fig:e1-e2} shows the large range of $e_1$ and $e_2$ consistent with our TTV/RV analysis. Note, however, that $e_1$ and $e_2$ cannot both be zero. Our analysis does not formally exclude high eccentricity solutions. These solutions, however, are disfavored for stability reasons and because TTV-active systems are observed to have eccentricities of a few percent.

Various groups have characterized the distribution of eccentricities among large numbers of \Kepler multi-planet systems, modeling eccentricities as a Rayleigh distribution parameterized by a mean eccentricity $\meanecc$. Studies of TTV-active multi-planet systems have found $\meanecc$ = 0.01--0.03 \citep{Wu13,Hadden14}. Analyses of transit durations in multi-planet systems where the host stars have well-measured densities have found \meanecc = 0.05--0.07 \citep{VanEylen15,Xie16}. That TTV-active systems exhibit lower \meanecc than the more general class of multi-planet systems suggests a distinct formation pathway.

Under the assumption that K2-24 is drawn from the population of TTV-active \Kepler multi-planet systems, we applied a Rayleigh prior on eccentricity \meanecc = 0.03. Figure~\ref{fig:e1-e2} shows the joint distribution of $e_1$ and $e_2$ including this prior. The eccentricity of K2-24c assumes the prior distribution. Solutions with non-zero $e_1$ are favored because $e_1 \sim 0$ requires $e_2 \sim 0.2$, which is strongly disfavored by our prior. For the remainder of the paper, we adopt $e_1$ = \lithwick{e1_fmt} and $e_2 < \lithwick{e2_p90}$ (90\% conf.). We discuss the dynamical origins of these eccentricities in Section~\ref{sec:dynamics}.

\section{Dynamics}
\label{sec:dynamics}
Here, we explore the dynamical origins of the K2-24 system architecture. In Section~\ref{sec:secular}, we discuss how the system evolves on secular timescales. In Section~\ref{sec:origin-eccentricity}, we consider several formation scenarios and assess whether they are consistent with the observed eccentricities.

\subsection{Secular Evolution}
\label{sec:secular}
While K2-24b and c are near the 2:1 mean-motion resonance, they cannot be locked in resonance. Resonant locking generally requires that $e \gtrsim \Delta^2 / \mu$, and for both planets $\Delta^2 / \mu \sim 3$. Therefore, the long-term dynamical evolution of K2-24b and c is dominated by secular interactions. The coplanar secular evolution of the planets' eccentricities may be visualized as trajectories in the $e$-$\Delta \varpi$ plane, where $\Delta \varpi$ is the angle between the apses.%
\footnote{Strictly speaking, the orbital angle relevant to the secular evolution is the longitude of perihelion $\varpi$ rather than the argument $\omega$. However, because we take the planetary orbits to be coplanar $\Delta \omega = \Delta \varpi$.}

\begin{figure*}
\centering
\includegraphics[trim=0.1cm 17cm 7.25cm 0cm, clip,width=1\textwidth]{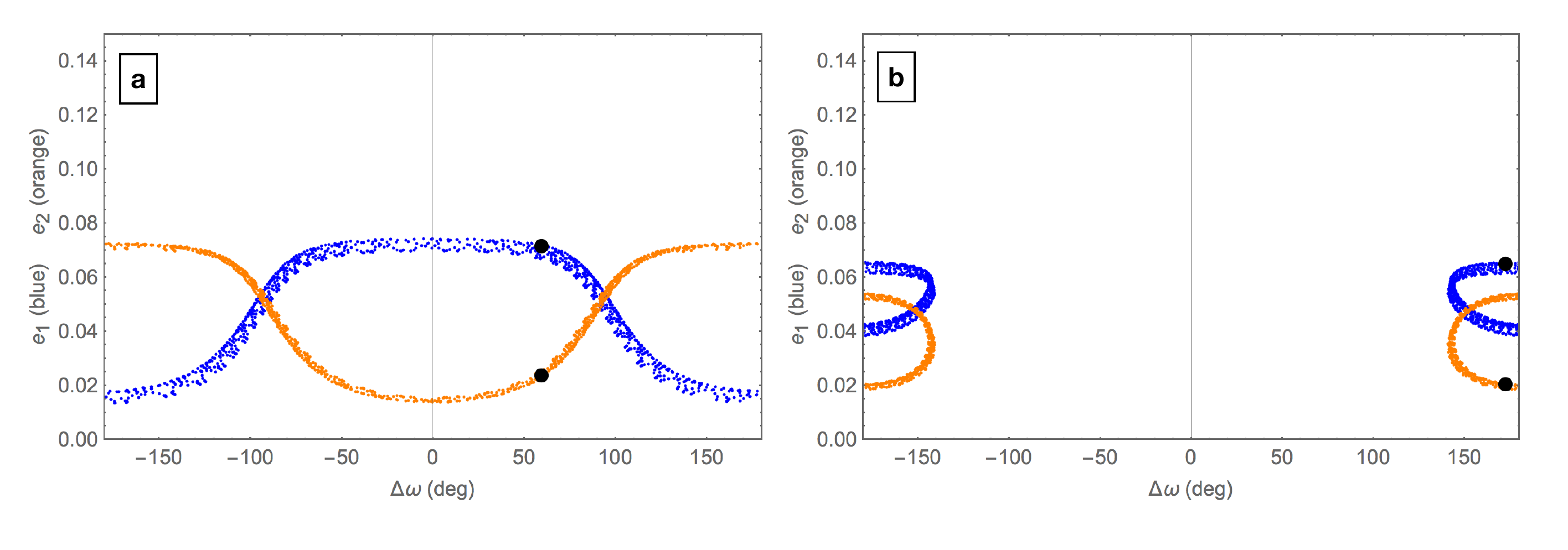}
\caption{Representative phase space trajectories for K2-24b (blue) and K2-24c (orange). {\em Left:} The x-axis shows the angle between the planet apses $\Delta \varpi$, and the y-axis shows the eccentricities. The dots show the starting values of the integration. In this realization, $\Delta \varpi$ circulates through all possible angles. {\em Right:} same except in this realization,  $\Delta \varpi$ librates about 180 deg (anti-alignment).\label{fig:secular}}
\end{figure*}

We simulated plausible long term evolutions of K2-24b and c by taking 1000 draws from the posterior samples from Section~\ref{sec:constraints} and integrating them for 10,000~years with the Mercury $N$-body integrator \citep{Chambers99}. These integrations revealed several qualitative apsidal outcomes: circulation, libration about $\Delta \varpi = 0^{\circ}$ (aligned apses), and libration about $\Delta \varpi = 180^{\circ}$ (anti-aligned apses). Indeed, the observational data is not yet precise enough to conclusively determine which of these regimes the systems actually occupies. We show representative examples of circulation and libration in Figure~\ref{fig:secular}. Inspection of these solutions shows that while at present time $\ecc{1}$ is likely larger than $\ecc{2}$, at other phases of the secular cycle \ecc{2} may be larger than \ecc{1}.

\subsection{Origin of Eccentricities}
\label{sec:origin-eccentricity}
Here, we consider several plausible mechanisms for exciting eccentricities, and assess whether they are consistent with the observed eccentricities of K2-24b and c. 

\subsubsection{Self-Excitation}
\label{sec:ecc-self}
We first considered the possibility that the eccentricities are self-excited, since gravitational interactions between two planets on initially circular orbits will pump eccentricities up to a certain value. To simulate this we performed an integration with Mercury using representative planet masses and setting the initial eccentricity to zero. As expected, the planets gained some eccentricity, but never exceeded $e = 0.005$. Eccentricities smaller than 0.005 are excluded by the data (see Figure~\ref{fig:e1-e2}), implying that some other process is required to explain the observed eccentricities.

\subsubsection{Divergent Migration Through Resonance}
\label{sec:ecc-divergent}
A well-known mechanism to excite eccentricity is divergent migration through mean-motion resonance. In this scenario planets begin interior to resonance with zero eccentricity. As shown in \cite{Batygin13}, migration through resonance corresponds with a separatrix crossing, after which the planets emerge with non-zero eccentricities and anti-aligned apses ($\Delta \varpi = 180^{\circ}$). As shown in \cite{Batygin15b}, the exited relic eccentricities are set by the planet-star mass ratios $\mu$ and initial eccentricities, which are usually assumed to be small.

In models of early Solar System evolution by \cite{Tsiganis05}, such a resonance crossing is used to trigger the onset of a transient dynamical instability. We note that divergent migration could be driven by gravitational scattering with a planetesimal disk \citep{Minton14}. 

Figure~\ref{fig:simulation} shows the time evolution of a simulation where K2-24b and c are adiabatically driven through resonance using fictitious forces. During the resonant crossing, eccentricities are quickly excited to \ecc{1} = 0.03 and \ecc{2} = 0.02. In this scenario, $\Delta \varpi$ is driven to 180~deg, and the libration amplitude is very small. Given that this mechanism produces planets that are stationary in the $e$--$\Delta \varpi$ plane, we can directly compare the present day $e$ to the predicted values from divergent migration.

In Figure~\ref{fig:e1-e2}, we compare the predicted eccentricities to our present day constraints. Eccentricities of $(e_1,e_2) = (0.03,0.02)$ are disfavored by the data, both with and without the Rayleigh prior on eccentricity. Moreover, the mechanism that drives divergent migration (e.g. planetesimal scattering) is also likely to damp eccentricities. Therefore, $(e_1,e_2) = (0.03,0.02)$ corresponds to upper bounds on the eccentricities the planets could acquire through this mechanism. This tension disfavors divergent resonant crossing as the sole explanation for the planet eccentricities, but future measurements of $\ecc{}$ and $\varpi$ for both planets would shed additional light on this interpretation.

\subsubsection{Disk-Driven Stochastic Excitation}
\label{sec:ecc-stochastic}

Another mechanism that excites eccentricities is stochastic interactions between young planets and a turbulent disk \citep{Adams08}. Density fluctuations within a turbulent protoplanetary disk cause eccentricities to grow approximately like a random walk, with $\mathrm{RMS}(e) \propto \sqrt{t}$. One mechanism to drive density fluctuations is the magnetorotational instability (MRI). In the limit of ideal MRI-driven turbulence, \cite{Okuzumi13} showed that the growth of $e$ can be constructed from analytical arguments:
\begin{eqnarray}
		\mathrm{RMS}(e)
        & \sim & 0.033
             \left(\frac{\alpha}{0.01}\right)^{1/2}
	         \left(\frac{\Sigma}{10^3\, \mathrm{g}\, \mathrm{cm}^2}\right)
	         \left(\frac{a}{0.1 \, \mathrm{AU}}\right)^{2} \nonumber\\ 
        & \times &
             \left(\frac{M_\star}{M_\odot}\right)^{-1}
             \left(\frac{n}{100\, \mathrm{days}^{-1}}\right)^{1/2}
             \left(\frac{t}{10\, \mathrm{Myr}}\right)^{1/2}
             \nonumber 
\end{eqnarray}
where $\alpha$ is Shakura-Sunyaev viscosity parameter, $\sigma$ is the surface density, and $n$ is the mean-motion. This equation suggests that if planets are embedded in a gas disk for a significant fraction of a 10~Myr disk lifetime, as they must have been to capture their H/He envelopes, they can acquire the several percent eccentricities we observe today.

\begin{figure*}
\centering
\includegraphics[trim=0cm 10cm 10.2cm 0.5cm, clip,width=1\textwidth]{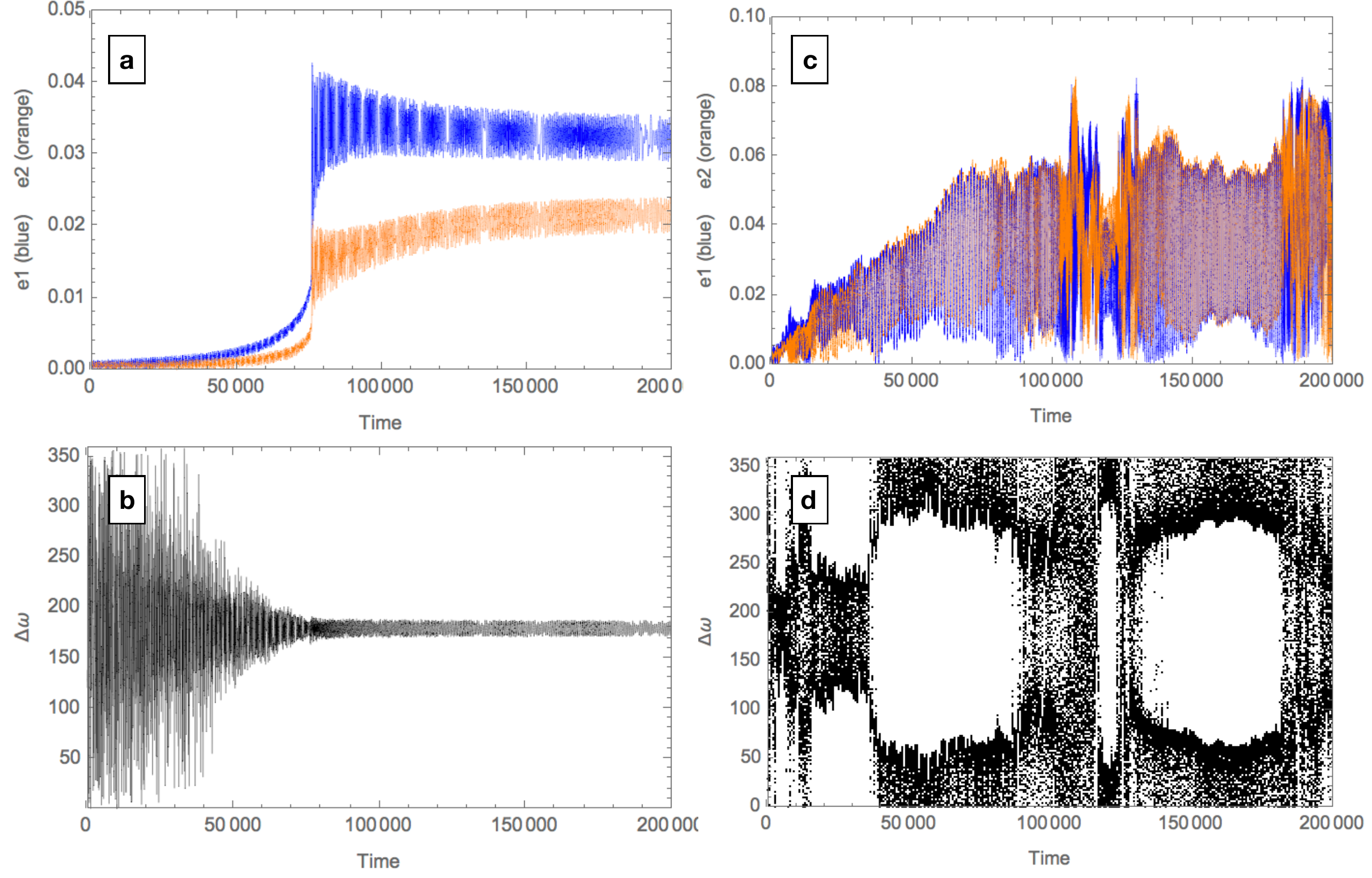}
\caption{Panels (a)--(b): Possible early time evolution of planet eccentricities and apsidal alignment angles as planets migrate divergently through the 2:1 resonance (see Section~\ref{sec:ecc-divergent}). Panel (a): During the resonance crossing the eccentricities are excited to \ecc{1} = 0.02 and \ecc{2} = 0.03. Panel (b): At early times, the orbits are nearly circular and $\Delta \varpi$ sweeps all angles between 0--360~deg. After the resonance crossing, the planets are anti-aligned with $\Delta \varpi$ = 180~deg. Panels (c)--(d): same as panels (a)--(b), but for planets subject to stochastic velocity perturbations (see Section~\ref{sec:ecc-stochastic}). Panel (c): eccentricities grow approximately like a random walk, with $\mathrm{RMS}(e) \propto \sqrt{t}$. Panel (d): There is no preferred value for $\Delta \varpi$.\label{fig:simulation}}
\end{figure*}

In order to illustrate this process, we performed a Mercury integration where we subjected the planets to appropriately scaled stochastic velocity kicks over a period of $2\times10^{5}$~yr. The simulation setup was identical to that of \cite{Batygin17}. The resulting evolution is shown in Figure~\ref{fig:simulation}. Note that unlike the case of divergent migration through resonance, the apsidal offset $\Delta \varpi$ takes on a broad range of values, resulting in an observable distinction between the two dynamical excitation mechanisms.

\subsubsection{Summary}
We considered three mechanisms for exciting planet eccentricities: self-excitation, divergent migration, and stochastic pumping. We found that self-excitation cannot explain the present day eccentricities. Divergent migration produces eccentricities that are qualitatively similar to the values observed today, although the predicted eccentricities are formally inconsistent with our measured values. Stochastic pumping can account for the present day eccentricities.

We stress that this is not an exhaustive analysis of excitation mechanisms. Among the mechanisms considered, however, stochastic pumping remains the most plausible explanation, given the data. Divergent migration predicts specific values for \ecc{1}, \ecc{2}, and $\Delta \varpi$ which can be corroborated with future observations. For example, measurements of secondary eclipse times place tight constraints on \ecosw{}. When combined with the constraints from this paper, such measurements would constrain $e$ and $\varpi$ separately. 

\section{Conclusions}
\label{sec:conclusions}
We have presented a joint TTV/RV analysis of the K2-24 system based on RVs from Keck/HIRES and transit observations with \ktwo and \spitzer. Our analysis provides new constraints on planet masses and core/envelope structure. Importantly, we leveraged the synergies between TTV and RV measurements to provide tight constraints on planet eccentricities of $e_1 \sim e_2 \sim 0.08$. Assuming the planets are drawn from the ensemble of \Kepler multi-planet systems, we found a small, but significantly non-zero eccentricity of \lithwick{e1_fmt} for K2-24b and we ruled out eccentricities larger than \lithwick{e2_p90} for K2-24c. These eccentricities are relics of the planets' past formation histories, and we found that stochastic interactions with a gas disk is a viable explanations for the observed dynamical state.

Future advances in the exoplanet census and RV instruments will expand the number of systems amenable to similar studies. Next-generation RV facilities at large telescopes such as VLT/ESPRESSO \citep{Gonzalez-Hernandez17}, Keck/KPF \citep{Gibson16}, and GMT/GCLEF \citep{Szentgyorgyi16} will enable RV measurements of a large sample of faint \Kepler planet hosts, including many TTV-active systems. Also, ESA's {\em PLATO} mission \citep{Rauer13} will conduct a transit survey over $\approx$2000~deg$^{2}$ for 2--3 years and add to the sample of planets with long baseline photometry.

Proceeding along an orthogonal direction, NASA's {\em TESS} mission \citep{Ricker14} will soon survey the entire sky, casting a wide net for planets around bright stars. These bright stars will be more amenable to RV follow-up than our current sample from \Kepler and \ktwo. One challenge is the limited baseline of {\em TESS} observations. During a nominal two-year mission, most of the sky would receive 27 days of {\em TESS} observations. While this will be sufficient to detect near-resonant systems, the baseline is too short to adequately sample TTV super-periods, which are typically measured in years. Extensions to {\em TESS} that would allow for subsequent transit measurements of known planets would therefore be exceedingly valuable.
 
\acknowledgements
We thank the anonymous reviewer for helpful suggestions that improved this manuscript. E.A.P. acknowledges support from Hubble Fellowship grant HST-HF2-51365.001-A awarded by the Space Telescope Science Institute, which is operated by the Association of Universities for Research in Astronomy, Inc., for NASA under contract NAS 5-26555. This work made use of NASA's Astrophysics Data System Bibliographic Services.

We thank Spitzer Science Center Director Tom Soifer for awarding discretionary time. This work is based in part on observations made with the Spitzer Space Telescope, which is operated by the Jet Propulsion Laboratory, California Institute of Technology under a contract with NASA. Support for this work was provided by NASA through an award issued by JPL/Caltech

Some of the data presented herein were obtained at the W.\ M.\ Keck Observatory, which is operated as a scientific partnership among the California Institute of Technology, the University of California, and NASA. We thank Andrew Howard, Lea Hirsch, Lauren Weiss, Howard Isaacson, and Molly Kosiarek for their assistance with the Keck/HIRES observations. The authors wish to recognize and acknowledge the very significant cultural role and reverence that the summit of Maunakea has always had within the indigenous Hawaiian community.  We are most fortunate to have the opportunity to conduct observations from this mountain.

\software{Astropy \citep{Astropy-Collaboration13}, batman \citep{Kreidberg15}, ChainConsumer \citep{Hinton2016}, emcee \citep{Foreman-Mackey13}, isoclassify \citep{Huber17}, lmfit \citep{Newville14}, matplotlib \citep{Hunter07}, Mercury \citep{Chambers99}, numpy/scipy \citep{numpy/scipy}, pandas \citep{pandas}, and RadVel \citep{Fulton18a}.}

\facilities{Keck:I (HIRES), Spitzer, Kepler}


\bibliography{manuscript.bib}

\begin{thebibliography}{}
\expandafter\ifx\csname natexlab\endcsname\relax\def\natexlab#1{#1}\fi

\bibitem[{{Adams} {et~al.}(2008){Adams}, {Laughlin}, \& {Bloch}}]{Adams08}
{Adams}, F.~C., {Laughlin}, G., \& {Bloch}, A.~M. 2008, \apj, 683, 1117

\bibitem[{{Agol} {et~al.}(2005){Agol}, {Steffen}, {Sari}, \&
  {Clarkson}}]{Agol05}
{Agol}, E., {Steffen}, J., {Sari}, R., \& {Clarkson}, W. 2005, \mnras, 359, 567

\bibitem[{{Astropy Collaboration} {et~al.}(2013){Astropy Collaboration},
  {Robitaille}, {Tollerud}, {Greenfield}, {Droettboom}, {Bray}, {Aldcroft},
  {Davis}, {Ginsburg}, {Price-Whelan}, {Kerzendorf}, {Conley}, {Crighton},
  {Barbary}, {Muna}, {Ferguson}, {Grollier}, {Parikh}, {Nair}, {Unther},
  {Deil}, {Woillez}, {Conseil}, {Kramer}, {Turner}, {Singer}, {Fox}, {Weaver},
  {Zabalza}, {Edwards}, {Azalee Bostroem}, {Burke}, {Casey}, {Crawford},
  {Dencheva}, {Ely}, {Jenness}, {Labrie}, {Lim}, {Pierfederici}, {Pontzen},
  {Ptak}, {Refsdal}, {Servillat}, \& {Streicher}}]{Astropy-Collaboration13}
{Astropy Collaboration}, {Robitaille}, T.~P., {Tollerud}, E.~J., {et~al.} 2013,
  \aap, 558, A33

\bibitem[{{Batygin}(2015)}]{Batygin15b}
{Batygin}, K. 2015, \mnras, 451, 2589

\bibitem[{{Batygin} \& {Adams}(2017)}]{Batygin17}
{Batygin}, K., \& {Adams}, F.~C. 2017, \aj, 153, 120

\bibitem[{{Batygin} \& {Morbidelli}(2013)}]{Batygin13}
{Batygin}, K., \& {Morbidelli}, A. 2013, \aap, 556, A28

\bibitem[{{Benneke} {et~al.}(2017){Benneke}, {Werner}, {Petigura}, {Knutson},
  {Dressing}, {Crossfield}, {Schlieder}, {Livingston}, {Beichman},
  {Christiansen}, {Krick}, {Gorjian}, {Howard}, {Sinukoff}, {Ciardi}, \&
  {Akeson}}]{Benneke17}
{Benneke}, B., {Werner}, M., {Petigura}, E., {et~al.} 2017, \apj, 834, 187

\bibitem[{{Borucki} {et~al.}(2010){Borucki}, {Koch}, {Basri}, {Batalha},
  {Brown}, {Caldwell}, {Caldwell}, {Christensen-Dalsgaard}, {Cochran},
  {DeVore}, {Dunham}, {Dupree}, {Gautier}, {Geary}, {Gilliland}, {Gould},
  {Howell}, {Jenkins}, {Kondo}, {Latham}, {Marcy}, {Meibom}, {Kjeldsen},
  {Lissauer}, {Monet}, {Morrison}, {Sasselov}, {Tarter}, {Boss}, {Brownlee},
  {Owen}, {Buzasi}, {Charbonneau}, {Doyle}, {Fortney}, {Ford}, {Holman},
  {Seager}, {Steffen}, {Welsh}, {Rowe}, {Anderson}, {Buchhave}, {Ciardi},
  {Walkowicz}, {Sherry}, {Horch}, {Isaacson}, {Everett}, {Fischer}, {Torres},
  {Johnson}, {Endl}, {MacQueen}, {Bryson}, {Dotson}, {Haas}, {Kolodziejczak},
  {Van Cleve}, {Chandrasekaran}, {Twicken}, {Quintana}, {Clarke}, {Allen},
  {Li}, {Wu}, {Tenenbaum}, {Verner}, {Bruhweiler}, {Barnes}, \&
  {Prsa}}]{Borucki10a}
{Borucki}, W.~J., {Koch}, D., {Basri}, G., {et~al.} 2010, Science, 327, 977

\bibitem[{{Brewer} {et~al.}(2016){Brewer}, {Fischer}, {Valenti}, \&
  {Piskunov}}]{Brewer16}
{Brewer}, J.~M., {Fischer}, D.~A., {Valenti}, J.~A., \& {Piskunov}, N. 2016,
  \apjs, 225, 32

\bibitem[{{Carter} {et~al.}(2012){Carter}, {Agol}, {Chaplin}, {Basu},
  {Bedding}, {Buchhave}, {Christensen-Dalsgaard}, {Deck}, {Elsworth},
  {Fabrycky}, {Ford}, {Fortney}, {Hale}, {Handberg}, {Hekker}, {Holman},
  {Huber}, {Karoff}, {Kawaler}, {Kjeldsen}, {Lissauer}, {Lopez}, {Lund},
  {Lundkvist}, {Metcalfe}, {Miglio}, {Rogers}, {Stello}, {Borucki}, {Bryson},
  {Christiansen}, {Cochran}, {Geary}, {Gilliland}, {Haas}, {Hall}, {Howard},
  {Jenkins}, {Klaus}, {Koch}, {Latham}, {MacQueen}, {Sasselov}, {Steffen},
  {Twicken}, \& {Winn}}]{Carter12}
{Carter}, J.~A., {Agol}, E., {Chaplin}, W.~J., {et~al.} 2012, Science, 337, 556

\bibitem[{{Chambers}(1999)}]{Chambers99}
{Chambers}, J.~E. 1999, \mnras, 304, 793

\bibitem[{{Deming} {et~al.}(2015){Deming}, {Knutson}, {Kammer}, {Fulton},
  {Ingalls}, {Carey}, {Burrows}, {Fortney}, {Todorov}, {Agol}, {Cowan},
  {Desert}, {Fraine}, {Langton}, {Morley}, \& {Showman}}]{Deming15}
{Deming}, D., {Knutson}, H., {Kammer}, J., {et~al.} 2015, \apj, 805, 132

\bibitem[{{Earl} \& {Deem}(2005)}]{Earl05}
{Earl}, D.~J., \& {Deem}, M.~W. 2005, Physical Chemistry Chemical Physics
  (Incorporating Faraday Transactions), 7, 3910

\bibitem[{{Eastman} {et~al.}(2013){Eastman}, {Gaudi}, \& {Agol}}]{Eastman13}
{Eastman}, J., {Gaudi}, B.~S., \& {Agol}, E. 2013, \pasp, 125, 83

\bibitem[{{Foreman-Mackey} {et~al.}(2013){Foreman-Mackey}, {Hogg}, {Lang}, \&
  {Goodman}}]{Foreman-Mackey13}
{Foreman-Mackey}, D., {Hogg}, D.~W., {Lang}, D., \& {Goodman}, J. 2013, \pasp,
  125, 306

\bibitem[{{Fulton} \& {Petigura}(2018)}]{Fulton18b}
{Fulton}, B.~J., \& {Petigura}, E.~A. 2018, ArXiv e-prints, arXiv:1805.01453

\bibitem[{{Fulton} {et~al.}(2018){Fulton}, {Petigura}, {Blunt}, \&
  {Sinukoff}}]{Fulton18a}
{Fulton}, B.~J., {Petigura}, E.~A., {Blunt}, S., \& {Sinukoff}, E. 2018, ArXiv
  e-prints, arXiv:1801.01947

\bibitem[{{Gaia Collaboration} {et~al.}(2018){Gaia Collaboration}, {Brown},
  {Vallenari}, {Prusti}, {de Bruijne}, {Babusiaux}, \& {Bailer-Jones}}]{Gaia18}
{Gaia Collaboration}, {Brown}, A.~G.~A., {Vallenari}, A., {et~al.} 2018, ArXiv
  e-prints, arXiv:1804.09365

\bibitem[{Gelman \& Rubin(1992)}]{Gelman92}
Gelman, A., \& Rubin, D.~B. 1992, Statistical Science, 7, 457

\bibitem[{{Gibson} {et~al.}(2016){Gibson}, {Howard}, {Marcy}, {Edelstein},
  {Wishnow}, \& {Poppett}}]{Gibson16}
{Gibson}, S.~R., {Howard}, A.~W., {Marcy}, G.~W., {et~al.} 2016, in \procspie,
  Vol. 9908, Ground-based and Airborne Instrumentation for Astronomy VI, 990870

\bibitem[{{Gonz{\'a}lez Hern{\'a}ndez} {et~al.}(2017){Gonz{\'a}lez
  Hern{\'a}ndez}, {Pepe}, {Molaro}, \& {Santos}}]{Gonzalez-Hernandez17}
{Gonz{\'a}lez Hern{\'a}ndez}, J.~I., {Pepe}, F., {Molaro}, P., \& {Santos}, N.
  2017, ArXiv e-prints, arXiv:1711.05250

\bibitem[{Goodman \& Weare(2010)}]{Goodman10}
Goodman, J., \& Weare, J. 2010, Communications in Applied Mathematics and
  Computational Science, 5, 65

\bibitem[{{Grillmair} {et~al.}(2012){Grillmair}, {Carey}, {Stauffer}, {Fisher},
  {Olds}, {Ingalls}, {Krick}, {Glaccum}, {Laine}, {Lowrance}, \&
  {Surace}}]{Grillmair12}
{Grillmair}, C.~J., {Carey}, S.~J., {Stauffer}, J.~R., {et~al.} 2012, in
  \procspie, Vol. 8448, Observatory Operations: Strategies, Processes, and
  Systems IV, 84481I

\bibitem[{{Hadden} \& {Lithwick}(2014)}]{Hadden14}
{Hadden}, S., \& {Lithwick}, Y. 2014, \apj, 787, 80

\bibitem[{{Hinton}(2016)}]{Hinton2016}
{Hinton}, S.~R. 2016, The Journal of Open Source Software, 1, 00045

\bibitem[{{Holczer} {et~al.}(2016){Holczer}, {Mazeh}, {Nachmani},
  {Jontof-Hutter}, {Ford}, {Fabrycky}, {Ragozzine}, {Kane}, \&
  {Steffen}}]{Holczer16}
{Holczer}, T., {Mazeh}, T., {Nachmani}, G., {et~al.} 2016, \apjs, 225, 9

\bibitem[{{Holman} \& {Murray}(2005)}]{Holman05}
{Holman}, M.~J., \& {Murray}, N.~W. 2005, Science, 307, 1288

\bibitem[{{Holman} {et~al.}(2010){Holman}, {Fabrycky}, {Ragozzine}, {Ford},
  {Steffen}, {Welsh}, {Lissauer}, {Latham}, {Marcy}, {Walkowicz}, {Batalha},
  {Jenkins}, {Rowe}, {Cochran}, {Fressin}, {Torres}, {Buchhave}, {Sasselov},
  {Borucki}, {Koch}, {Basri}, {Brown}, {Caldwell}, {Charbonneau}, {Dunham},
  {Gautier}, {Geary}, {Gilliland}, {Haas}, {Howell}, {Ciardi}, {Endl},
  {Fischer}, {F{\"u}r{\'e}sz}, {Hartman}, {Isaacson}, {Johnson}, {MacQueen},
  {Moorhead}, {Morehead}, \& {Orosz}}]{Holman10}
{Holman}, M.~J., {Fabrycky}, D.~C., {Ragozzine}, D., {et~al.} 2010, Science,
  330, 51

\bibitem[{{Howard} \& {Fulton}(2016)}]{Howard16}
{Howard}, A.~W., \& {Fulton}, B.~J. 2016, \pasp, 128, 114401

\bibitem[{{Howard} {et~al.}(2010){Howard}, {Johnson}, {Marcy}, {Fischer},
  {Wright}, {Bernat}, {Henry}, {Peek}, {Isaacson}, {Apps}, {Endl}, {Cochran},
  {Valenti}, {Anderson}, \& {Piskunov}}]{Howard10b}
{Howard}, A.~W., {Johnson}, J.~A., {Marcy}, G.~W., {et~al.} 2010, \apj, 721,
  1467

\bibitem[{{Howard} {et~al.}(2013){Howard}, {Sanchis-Ojeda}, {Marcy}, {Johnson},
  {Winn}, {Isaacson}, {Fischer}, {Fulton}, {Sinukoff}, \& {Fortney}}]{Howard13}
{Howard}, A.~W., {Sanchis-Ojeda}, R., {Marcy}, G.~W., {et~al.} 2013, \nat, 503,
  381

\bibitem[{{Howell} {et~al.}(2014){Howell}, {Sobeck}, {Haas}, {Still},
  {Barclay}, {Mullally}, {Troeltzsch}, {Aigrain}, {Bryson}, {Caldwell},
  {Chaplin}, {Cochran}, {Huber}, {Marcy}, {Miglio}, {Najita}, {Smith},
  {Twicken}, \& {Fortney}}]{Howell14}
{Howell}, S.~B., {Sobeck}, C., {Haas}, M., {et~al.} 2014, \pasp, 126, 398

\bibitem[{{Huber} {et~al.}(2017){Huber}, {Zinn}, {Bojsen-Hansen},
  {Pinsonneault}, {Sahlholdt}, {Serenelli}, {Silva Aguirre}, {Stassun},
  {Stello}, {Tayar}, {Bastien}, {Bedding}, {Buchhave}, {Chaplin}, {Davies},
  {Garc{\'{\i}}a}, {Latham}, {Mathur}, {Mosser}, \& {Sharma}}]{Huber17}
{Huber}, D., {Zinn}, J., {Bojsen-Hansen}, M., {et~al.} 2017, \apj, 844, 102

\bibitem[{Hunter(2007)}]{Hunter07}
Hunter, J.~D. 2007, Computing In Science \& Engineering, 9, 90

\bibitem[{{Ingalls} {et~al.}(2012){Ingalls}, {Krick}, {Carey}, {Laine},
  {Surace}, {Glaccum}, {Grillmair}, \& {Lowrance}}]{Ingalls12}
{Ingalls}, J.~G., {Krick}, J.~E., {Carey}, S.~J., {et~al.} 2012, in \procspie,
  Vol. 8442, Space Telescopes and Instrumentation 2012: Optical, Infrared, and
  Millimeter Wave, 84421Y

\bibitem[{{Kreidberg}(2015)}]{Kreidberg15}
{Kreidberg}, L. 2015, ArXiv e-prints, arXiv:1507.08285

\bibitem[{{Lithwick} {et~al.}(2012){Lithwick}, {Xie}, \& {Wu}}]{Lithwick12}
{Lithwick}, Y., {Xie}, J., \& {Wu}, Y. 2012, \apj, 761, 122

\bibitem[{{Lopez} \& {Fortney}(2014)}]{Lopez14}
{Lopez}, E.~D., \& {Fortney}, J.~J. 2014, \apj, 792, 1

\bibitem[{{Marcy} \& {Butler}(1992)}]{Marcy92}
{Marcy}, G.~W., \& {Butler}, R.~P. 1992, \pasp, 104, 270

\bibitem[{{Marcy} {et~al.}(2014){Marcy}, {Isaacson}, {Howard}, {Rowe},
  {Jenkins}, {Bryson}, {Latham}, {Howell}, {Gautier}, {Batalha}, {Rogers},
  {Ciardi}, {Fischer}, {Gilliland}, {Kjeldsen}, {Christensen-Dalsgaard},
  {Huber}, {Chaplin}, {Basu}, {Buchhave}, {Quinn}, {Borucki}, {Koch}, {Hunter},
  {Caldwell}, {Van Cleve}, {Kolbl}, {Weiss}, {Petigura}, {Seager}, {Morton},
  {Johnson}, {Ballard}, {Burke}, {Cochran}, {Endl}, {MacQueen}, {Everett},
  {Lissauer}, {Ford}, {Torres}, {Fressin}, {Brown}, {Steffen}, {Charbonneau},
  {Basri}, {Sasselov}, {Winn}, {Sanchis-Ojeda}, {Christiansen}, {Adams},
  {Henze}, {Dupree}, {Fabrycky}, {Fortney}, {Tarter}, {Holman}, {Tenenbaum},
  {Shporer}, {Lucas}, {Welsh}, {Orosz}, {Bedding}, {Campante}, {Davies},
  {Elsworth}, {Handberg}, {Hekker}, {Karoff}, {Kawaler}, {Lund}, {Lundkvist},
  {Metcalfe}, {Miglio}, {Silva Aguirre}, {Stello}, {White}, {Boss}, {Devore},
  {Gould}, {Prsa}, {Agol}, {Barclay}, {Coughlin}, {Brugamyer}, {Mullally},
  {Quintana}, {Still}, {Thompson}, {Morrison}, {Twicken}, {D{\'e}sert},
  {Carter}, {Crepp}, {H{\'e}brard}, {Santerne}, {Moutou}, {Sobeck}, {Hudgins},
  {Haas}, {Robertson}, {Lillo-Box}, \& {Barrado}}]{Marcy14}
{Marcy}, G.~W., {Isaacson}, H., {Howard}, A.~W., {et~al.} 2014, \apjs, 210, 20

\bibitem[{McKinney(2010)}]{pandas}
McKinney, W. 2010, in Proceedings of the 9th Python in Science Conference, ed.
  S.~van~der Walt \& J.~Millman, 51 -- 56

\bibitem[{{Millholland} {et~al.}(2018){Millholland}, {Laughlin}, {Teske},
  {Butler}, {Burt}, {Holden}, {Vogt}, {Crane}, {Shectman}, \&
  {Thompson}}]{Millholland18}
{Millholland}, S., {Laughlin}, G., {Teske}, J., {et~al.} 2018, ArXiv e-prints,
  arXiv:1801.07831

\bibitem[{{Mills} {et~al.}(2016){Mills}, {Fabrycky}, {Migaszewski}, {Ford},
  {Petigura}, \& {Isaacson}}]{Mills16}
{Mills}, S.~M., {Fabrycky}, D.~C., {Migaszewski}, C., {et~al.} 2016, \nat, 533,
  509

\bibitem[{{Mills} \& {Mazeh}(2017)}]{Mills17}
{Mills}, S.~M., \& {Mazeh}, T. 2017, \apjl, 839, L8

\bibitem[{{Minton} \& {Levison}(2014)}]{Minton14}
{Minton}, D.~A., \& {Levison}, H.~F. 2014, \icarus, 232, 118

\bibitem[{{Montet} {et~al.}(2015){Montet}, {Morton}, {Foreman-Mackey},
  {Johnson}, {Hogg}, {Bowler}, {Latham}, {Bieryla}, \& {Mann}}]{Montet15}
{Montet}, B.~T., {Morton}, T.~D., {Foreman-Mackey}, D., {et~al.} 2015, \apj,
  809, 25

\bibitem[{{Nelson} {et~al.}(2016){Nelson}, {Robertson}, {Payne}, {Pritchard},
  {Deck}, {Ford}, {Wright}, \& {Isaacson}}]{Nelson16}
{Nelson}, B.~E., {Robertson}, P.~M., {Payne}, M.~J., {et~al.} 2016, \mnras,
  455, 2484

\bibitem[{{Nesvorn{\'y}} {et~al.}(2013){Nesvorn{\'y}}, {Kipping}, {Terrell},
  {Hartman}, {Bakos}, \& {Buchhave}}]{Nesvorny13}
{Nesvorn{\'y}}, D., {Kipping}, D., {Terrell}, D., {et~al.} 2013, \apj, 777, 3

\bibitem[{Newville {et~al.}(2014)Newville, Stensitzki, Allen, \&
  Ingargiola}]{Newville14}
Newville, M., Stensitzki, T., Allen, D.~B., \& Ingargiola, A. 2014, LMFIT:
  Non-Linear Least-Square Minimization and Curve-Fitting for Python, , ,
  doi:10.5281/zenodo.11813

\bibitem[{{Okuzumi} \& {Ormel}(2013)}]{Okuzumi13}
{Okuzumi}, S., \& {Ormel}, C.~W. 2013, \apj, 771, 43

\bibitem[{{O'Toole} {et~al.}(2009){O'Toole}, {Jones}, {Tinney}, {Butler},
  {Marcy}, {Carter}, {Bailey}, \& {Wittenmyer}}]{Otoole09}
{O'Toole}, S.~J., {Jones}, H.~R.~A., {Tinney}, C.~G., {et~al.} 2009, \apj, 701,
  1732

\bibitem[{{Pepe} {et~al.}(2013){Pepe}, {Cameron}, {Latham}, {Molinari}, {Udry},
  {Bonomo}, {Buchhave}, {Charbonneau}, {Cosentino}, {Dressing}, {Dumusque},
  {Figueira}, {Fiorenzano}, {Gettel}, {Harutyunyan}, {Haywood}, {Horne},
  {Lopez-Morales}, {Lovis}, {Malavolta}, {Mayor}, {Micela}, {Motalebi},
  {Nascimbeni}, {Phillips}, {Piotto}, {Pollacco}, {Queloz}, {Rice}, {Sasselov},
  {S{\'e}gransan}, {Sozzetti}, {Szentgyorgyi}, \& {Watson}}]{Pepe13}
{Pepe}, F., {Cameron}, A.~C., {Latham}, D.~W., {et~al.} 2013, \nat, 503, 377

\bibitem[{{Petigura} {et~al.}(2016){Petigura}, {Howard}, {Lopez}, {Deck},
  {Fulton}, {Crossfield}, {Ciardi}, {Chiang}, {Lee}, {Isaacson}, {Beichman},
  {Hansen}, {Schlieder}, \& {Sinukoff}}]{Petigura16}
{Petigura}, E.~A., {Howard}, A.~W., {Lopez}, E.~D., {et~al.} 2016, \apj, 818,
  36

\bibitem[{{Petigura} {et~al.}(2017){Petigura}, {Sinukoff}, {Lopez},
  {Crossfield}, {Howard}, {Brewer}, {Fulton}, {Isaacson}, {Ciardi}, {Howell},
  {Everett}, {Horch}, {Hirsch}, {Weiss}, \& {Schlieder}}]{Petigura17a}
{Petigura}, E.~A., {Sinukoff}, E., {Lopez}, E.~D., {et~al.} 2017, \aj, 153, 142

\bibitem[{{Petigura} {et~al.}(2018){Petigura}, {Crossfield}, {Isaacson},
  {Beichman}, {Christiansen}, {Dressing}, {Fulton}, {Howard}, {Kosiarek},
  {L{\'e}pine}, {Schlieder}, {Sinukoff}, \& {Yee}}]{Petigura18}
{Petigura}, E.~A., {Crossfield}, I.~J.~M., {Isaacson}, H., {et~al.} 2018, \aj,
  155, 21

\bibitem[{{Pollack} {et~al.}(1996){Pollack}, {Hubickyj}, {Bodenheimer},
  {Lissauer}, {Podolak}, \& {Greenzweig}}]{Pollack96}
{Pollack}, J.~B., {Hubickyj}, O., {Bodenheimer}, P., {et~al.} 1996, Icarus,
  124, 62

\bibitem[{{Rauer}(2013)}]{Rauer13}
{Rauer}, H. 2013, European Planetary Science Congress, 8, EPSC2013

\bibitem[{{Ricker} {et~al.}(2014){Ricker}, {Winn}, {Vanderspek}, {Latham},
  {Bakos}, {Bean}, {Berta-Thompson}, {Brown}, {Buchhave}, {Butler}, {Butler},
  {Chaplin}, {Charbonneau}, {Christensen-Dalsgaard}, {Clampin}, {Deming},
  {Doty}, {De Lee}, {Dressing}, {Dunham}, {Endl}, {Fressin}, {Ge}, {Henning},
  {Holman}, {Howard}, {Ida}, {Jenkins}, {Jernigan}, {Johnson}, {Kaltenegger},
  {Kawai}, {Kjeldsen}, {Laughlin}, {Levine}, {Lin}, {Lissauer}, {MacQueen},
  {Marcy}, {McCullough}, {Morton}, {Narita}, {Paegert}, {Palle}, {Pepe},
  {Pepper}, {Quirrenbach}, {Rinehart}, {Sasselov}, {Sato}, {Seager},
  {Sozzetti}, {Stassun}, {Sullivan}, {Szentgyorgyi}, {Torres}, {Udry}, \&
  {Villasenor}}]{Ricker14}
{Ricker}, G.~R., {Winn}, J.~N., {Vanderspek}, R., {et~al.} 2014, in \procspie,
  Vol. 9143, Space Telescopes and Instrumentation 2014: Optical, Infrared, and
  Millimeter Wave, 914320

\bibitem[{{Rivera} {et~al.}(2010){Rivera}, {Laughlin}, {Butler}, {Vogt},
  {Haghighipour}, \& {Meschiari}}]{Rivera10}
{Rivera}, E.~J., {Laughlin}, G., {Butler}, R.~P., {et~al.} 2010, \apj, 719, 890

\bibitem[{{Rogers}(2015)}]{Rogers15}
{Rogers}, L.~A. 2015, \apj, 801, 41

\bibitem[{{Rowe} {et~al.}(2014){Rowe}, {Bryson}, {Marcy}, {Lissauer},
  {Jontof-Hutter}, {Mullally}, {Gilliland}, {Issacson}, {Ford}, {Howell},
  {Borucki}, {Haas}, {Huber}, {Steffen}, {Thompson}, {Quintana}, {Barclay},
  {Still}, {Fortney}, {Gautier}, {Hunter}, {Caldwell}, {Ciardi}, {Devore},
  {Cochran}, {Jenkins}, {Agol}, {Carter}, \& {Geary}}]{Rowe14}
{Rowe}, J.~F., {Bryson}, S.~T., {Marcy}, G.~W., {et~al.} 2014, \apj, 784, 45

\bibitem[{Schwarz(1978)}]{Schwartz78}
Schwarz, G. 1978, Annals of Statistics, 6, 461

\bibitem[{{Skrutskie} {et~al.}(2006){Skrutskie}, {Cutri}, {Stiening},
  {Weinberg}, {Schneider}, {Carpenter}, {Beichman}, {Capps}, {Chester},
  {Elias}, {Huchra}, {Liebert}, {Lonsdale}, {Monet}, {Price}, {Seitzer},
  {Jarrett}, {Kirkpatrick}, {Gizis}, {Howard}, {Evans}, {Fowler}, {Fullmer},
  {Hurt}, {Light}, {Kopan}, {Marsh}, {McCallon}, {Tam}, {Van Dyk}, \&
  {Wheelock}}]{Skrutskie06}
{Skrutskie}, M.~F., {Cutri}, R.~M., {Stiening}, R., {et~al.} 2006, \aj, 131,
  1163

\bibitem[{{Szentgyorgyi} {et~al.}(2016){Szentgyorgyi}, {Baldwin}, {Barnes},
  {Bean}, {Ben-Ami}, {Brennan}, {Budynkiewicz}, {Chun}, {Conroy}, {Crane},
  {Epps}, {Evans}, {Evans}, {Foster}, {Frebel}, {Gauron}, {Guzm{\'a}n}, {Hare},
  {Jang}, {Jang}, {Jordan}, {Kim}, {Kim}, {Mendes de Oliveira},
  {Lopez-Morales}, {McCracken}, {McMuldroch}, {Miller}, {Mueller}, {Oh},
  {Onyuksel}, {Ordway}, {Park}, {Park}, {Park}, {Paxson}, {Phillips},
  {Plummer}, {Podgorski}, {Seifahrt}, {Stark}, {Steiner}, {Uomoto},
  {Walsworth}, \& {Yu}}]{Szentgyorgyi16}
{Szentgyorgyi}, A., {Baldwin}, D., {Barnes}, S., {et~al.} 2016, in \procspie,
  Vol. 9908, Ground-based and Airborne Instrumentation for Astronomy VI, 990822

\bibitem[{{Tsiganis} {et~al.}(2005){Tsiganis}, {Gomes}, {Morbidelli}, \&
  {Levison}}]{Tsiganis05}
{Tsiganis}, K., {Gomes}, R., {Morbidelli}, A., \& {Levison}, H.~F. 2005, \nat,
  435, 459

\bibitem[{{Valenti} {et~al.}(1995){Valenti}, {Butler}, \& {Marcy}}]{Valenti95}
{Valenti}, J.~A., {Butler}, R.~P., \& {Marcy}, G.~W. 1995, \pasp, 107, 966

\bibitem[{van~der Walt {et~al.}(2011)van~der Walt, Colbert, \&
  Varoquaux}]{numpy/scipy}
van~der Walt, S., Colbert, S.~C., \& Varoquaux, G. 2011, Computing in Science
  Engineering, 13, 22

\bibitem[{{Van Eylen} \& {Albrecht}(2015)}]{VanEylen15}
{Van Eylen}, V., \& {Albrecht}, S. 2015, \apj, 808, 126

\bibitem[{{Vaughan} {et~al.}(1978){Vaughan}, {Preston}, \&
  {Wilson}}]{Vaughan78}
{Vaughan}, A.~H., {Preston}, G.~W., \& {Wilson}, O.~C. 1978, \pasp, 90, 267

\bibitem[{{Vogt} {et~al.}(1994){Vogt}, {Allen}, {Bigelow}, {Bresee}, {Brown},
  {Cantrall}, {Conrad}, {Couture}, {Delaney}, {Epps}, {Hilyard}, {Hilyard},
  {Horn}, {Jern}, {Kanto}, {Keane}, {Kibrick}, {Lewis}, {Osborne},
  {Pardeilhan}, {Pfister}, {Ricketts}, {Robinson}, {Stover}, {Tucker}, {Ward},
  \& {Wei}}]{Vogt94}
{Vogt}, S.~S., {Allen}, S.~L., {Bigelow}, B.~C., {et~al.} 1994, 2198, 362

\bibitem[{{Weiss} \& {Marcy}(2014)}]{Weiss14}
{Weiss}, L.~M., \& {Marcy}, G.~W. 2014, \apjl, 783, L6

\bibitem[{{Wu} \& {Lithwick}(2013)}]{Wu13}
{Wu}, Y., \& {Lithwick}, Y. 2013, \apj, 772, 74

\bibitem[{{Xie} {et~al.}(2016){Xie}, {Dong}, {Zhu}, {Huber}, {Zheng}, {De Cat},
  {Fu}, {Liu}, {Luo}, {Wu}, {Zhang}, {Zhang}, {Zhou}, {Cao}, {Hou}, {Wang}, \&
  {Zhang}}]{Xie16}
{Xie}, J.-W., {Dong}, S., {Zhu}, Z., {et~al.} 2016, Proceedings of the National
  Academy of Science, 113, 11431

\end{thebibliography}

\clearpage
\appendix

\section{TTV Modeling}
\label{sec:appendix}
Table~\ref{tab:transit-times-predict} lists the predicted transit times and uncertainties for K2-24b and c up to 2025.

\begin{deluxetable}{llrrr}
\tablecaption{Predicted Transit Times\label{tab:transit-times-predict}}
\tablehead{
  \colhead{Planet} & 
  \colhead{$i$} & 
  \colhead{UTC date} & 
  \colhead{$T_{c}$} & 
  \colhead{$\sigma(T_c)$} \\ 
  \colhead{} & 
  \colhead{} & 
  \colhead{} & 
  \colhead{days} & 
  \colhead{days} 
}
\startdata
b & 0 & 2014-09-05 & 2072.7962 & 0.0007 \\
c & 0 & 2014-09-15 & 2082.6250 & 0.0006 \\
b & 1 & 2014-09-26 & 2093.6803 & 0.0006 \\
b & 2 & 2014-10-17 & 2114.5650 & 0.0005 \\
c & 1 & 2014-10-27 & 2124.9877 & 0.0006 \\
\hline
b & 195 & 2025-10-31 & 6146.5011 & 0.0556 \\
c & 96 & 2025-10-31 & 6146.7839 & 0.1006 \\
b & 196 & 2025-11-20 & 6167.3924 & 0.0567 \\
b & 197 & 2025-12-11 & 6188.2829 & 0.0578 \\
c & 97 & 2025-12-12 & 6189.1158 & 0.1048 \\
\enddata
\tablecomments{Predicted transit times for K2-24b and c, where $i$, is an index that labels individual transits. Times are given in $\mathrm{BJD}_\mathrm{TBD} - 2454833$. Table~\ref{tab:transit-times} is published in its entirety in the machine-readable format. A portion is shown here for guidance regarding its form and content.}
\end{deluxetable}

\end{document}